\documentclass[10pt,prl,aps,showpacs,twocolumn,preprintnumbers,amsmath,amssymb,superscriptaddress]{revtex4-2}

\usepackage{graphicx}    
\usepackage{latexsym}    %
\usepackage{color}       %
\usepackage{dcolumn}     
\usepackage{bm}          
\usepackage{amssymb}     
\usepackage{hyperref}    
\expandafter\let\csname equation*\endcsname\relax
\expandafter\let\csname endequation*\endcsname\relax
\usepackage{amsmath}
\usepackage{mathtools}          

\usepackage{graphics}
\usepackage{epsfig}
\usepackage{subfigure}
\usepackage{nicefrac}
\usepackage{verbatim}
\usepackage{enumerate}
\usepackage{footnote}
\usepackage{booktabs}
\usepackage{siunitx}
\usepackage{xcolor}
\renewcommand{\arraystretch}{1.8}

\begin{document}

\title{Stochastic interpolation of sparsely sampled time series via multi-point fractional Brownian bridges}
\author{J.~Friedrich}
\affiliation{Univ. Lyon, ENS de Lyon, Univ. Claude Bernard, CNRS, Laboratoire de Physique,
F-69342, Lyon, France}
\author{S.~Gallon}
\affiliation{Univ. Lyon, ENS de Lyon, Univ. Claude Bernard, CNRS, Laboratoire de Physique,
F-69342, Lyon, France}
\affiliation{Institute for Theoretical Physics I, Ruhr-University Bochum, Universit\"{a}tsstr. 150,
D-44801 Bochum, Germany}
\author{A.~Pumir}
\affiliation{Univ. Lyon, ENS de Lyon, Univ. Claude Bernard, CNRS, Laboratoire de Physique,
F-69342, Lyon, France}
\author{R.~Grauer}
\affiliation{Institute for Theoretical Physics I, Ruhr-University Bochum, Universit\"{a}tsstr. 150,
D-44801 Bochum, Germany}
\date{\today}
\begin{abstract}
We propose and test a method to interpolate sparsely sampled signals by a stochastic process with a broad range of spatial and/or temporal scales. To this end, we extend the notion of a fractional Brownian bridge, defined as fractional Brownian motion with a given scaling (Hurst) exponent $H$ and with prescribed start and end points, to a bridge process with an arbitrary number of intermediate and non-equidistant points. Determining the optimal  value of the Hurst exponent, $H_{opt}$, appropriate to interpolate the sparse signal, is a very important step of our method. We demonstrate the validity of our method on a signal from fluid turbulence in a high Reynolds number flow and discuss the implications of the non-self-similar character of the signal. The method introduced here could be instrumental in several physical problems, including
astrophysics, particle tracking, specific tailoring of surrogate data,
as well as in domains of natural and social sciences.
\end{abstract}
\maketitle                             
Many non-equilibrium phenomena in physics involve
random fluctuations with a wide range of spatial and/or temporal scales~\cite{haken1983,prigogineeltit}.
The theory of stochastic processes provides a conceptual framework to
describe such phenomena~\cite{Wax,Friedrich2011a}.
The most emblematic example is provided by Brownian motion, which
results from random uncorrelated collisions acting on a particle, and
can be described by a Wiener process~\cite{Einstein1905}. Nonetheless, several complex systems in
nature~\cite{Kolmogorov1941,Onsager1949,Weizsacker1948,Goldstein1995,Makarava2014,Peng1993} also involve long- or short-range correlations which can be described in
terms of fractional Brownian motion~\cite{levy1965processus,mandelbrot1968}.
Furthermore, the description of 
a broad range of problems in physics or in the general field of complex systems requires the fractional Brownian motion (fBm)
process to be \emph{constrained}, by passing through a certain number of prescribed points.
Typical examples include the tracking of animal movement~\cite{kranstauber2012}, information-based financial models~\cite{Brody2008}, the number of neutrons involved in reactor diffusion~\cite{de_Mulatier_2015,Mazzolo2017}, or cosmic ray propagation in astrophysical turbulence~\cite{batista-etal:2016}.

Current stochastic interpolation  methods, e.g.,
kriging~\cite{cressie1990origins,Delhomme_1978}, which is used for data reconstruction/spatial enhancement of particle image velocimetry~\cite{doi:10.1063/1.3003069}, interpolations of buoy location in the tropical pacific~\cite{hansen1989temporal},
and interpolation of rainfall data from rain gages and radars~\cite{seo1990stochastic,seo1990stochastic1},
or other various types of polynomial interpolation act often to smooth the underlying data~\cite{griffa2004predictability}.
A more promising method to interpolate  sparse data sets
has been introduced in
the field of surface hydrology~\cite{molz1993fractal,molz1997fractional},
using the L\'evy-Ciesielski
construction~\cite{levy1965processus,ciesielski1961holder}. This construction,
however, is exact only in the very special case when the random processes are
describable by
ordinary Brownian motion~\cite{saupe1988algorithms,Voss_1988}.

Providing a more faithful representation of the data requires the development of exact stochastic interpolations on the basis of fBm while at the same time giving an estimate for the optimal Hurst exponent which characterizes the roughness of the sparsely sampled data.

Here, we stress the important example of cosmic ray propagation in turbulent magnetic fields~\cite{schlickeiser:2015, zweibel:2013}. In order to overcome the overwhelming problem of resolving the wide range of scales involved in these extremely high Reynolds number astrophysical flows,
several methods for generating synthetic turbulent fields were developed in the last decades~\cite{giacalone-jokipii:1999, zimbardo-amato-etal:2015,
snodin-shukurov-etal:2016,reichherzer-etal:2019}. Such methods are frequently  implemented and used in major cosmic ray propagation codes (see e.g.~\cite{batista-etal:2016}). To capture large anisotropies due to the geometry of galaxies (spiral arms, outflow regions, bow shocks), synthetic turbulent fields must be embedded in large-scale magnetohydrodynamic (MHD) simulations of the turbulent interstellar or intergalactic plasma. Therefore, in this problem we face two challenges: i.)\;fBm, which is used to emulate turbulent flow properties, must be constrained to match the values on the numerical grid of the MHD simulation and ii.)\;scaling properties, represented by the Hurst exponent of fluctuations of the coarse-grained MHD simulation must be determined
from sparse grid data, in order to allow for an ``optimal stochastic interpolation'' of sparsely-sampled
data. 

\begin{figure}[h!]
\centering
\includegraphics[width=0.4 \textwidth]{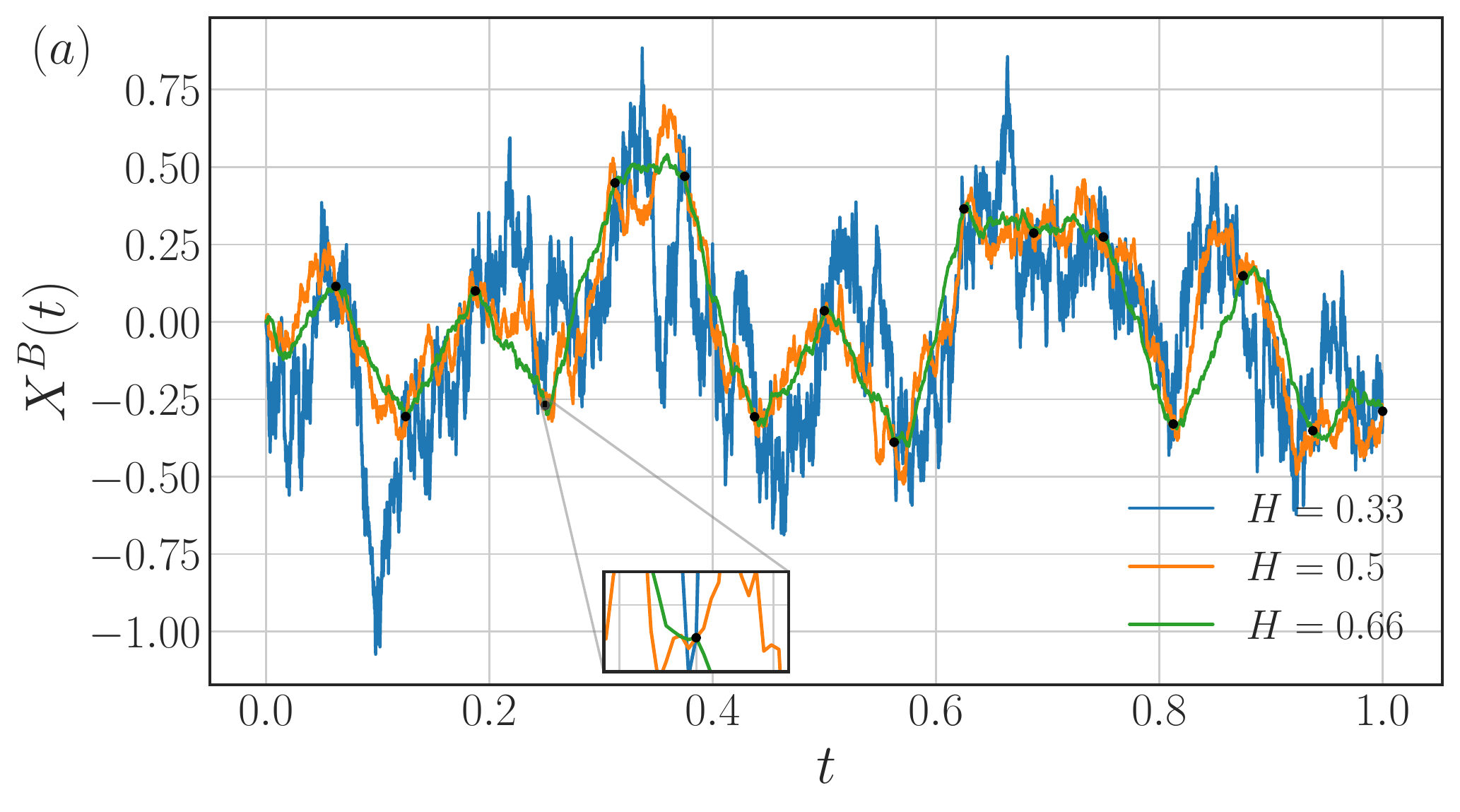}
\includegraphics[width=0.4 \textwidth]{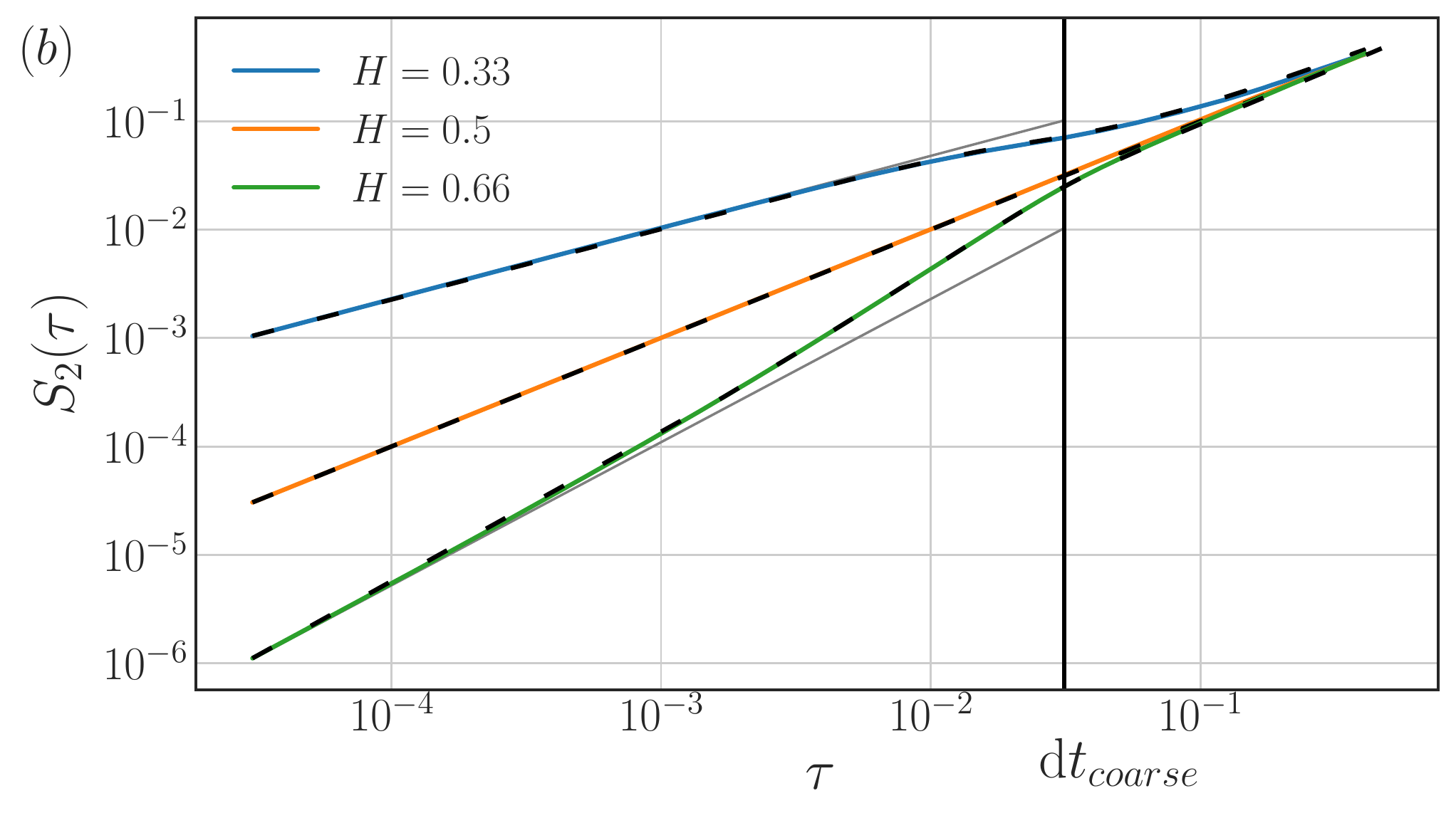}
\includegraphics[width=0.4 \textwidth]{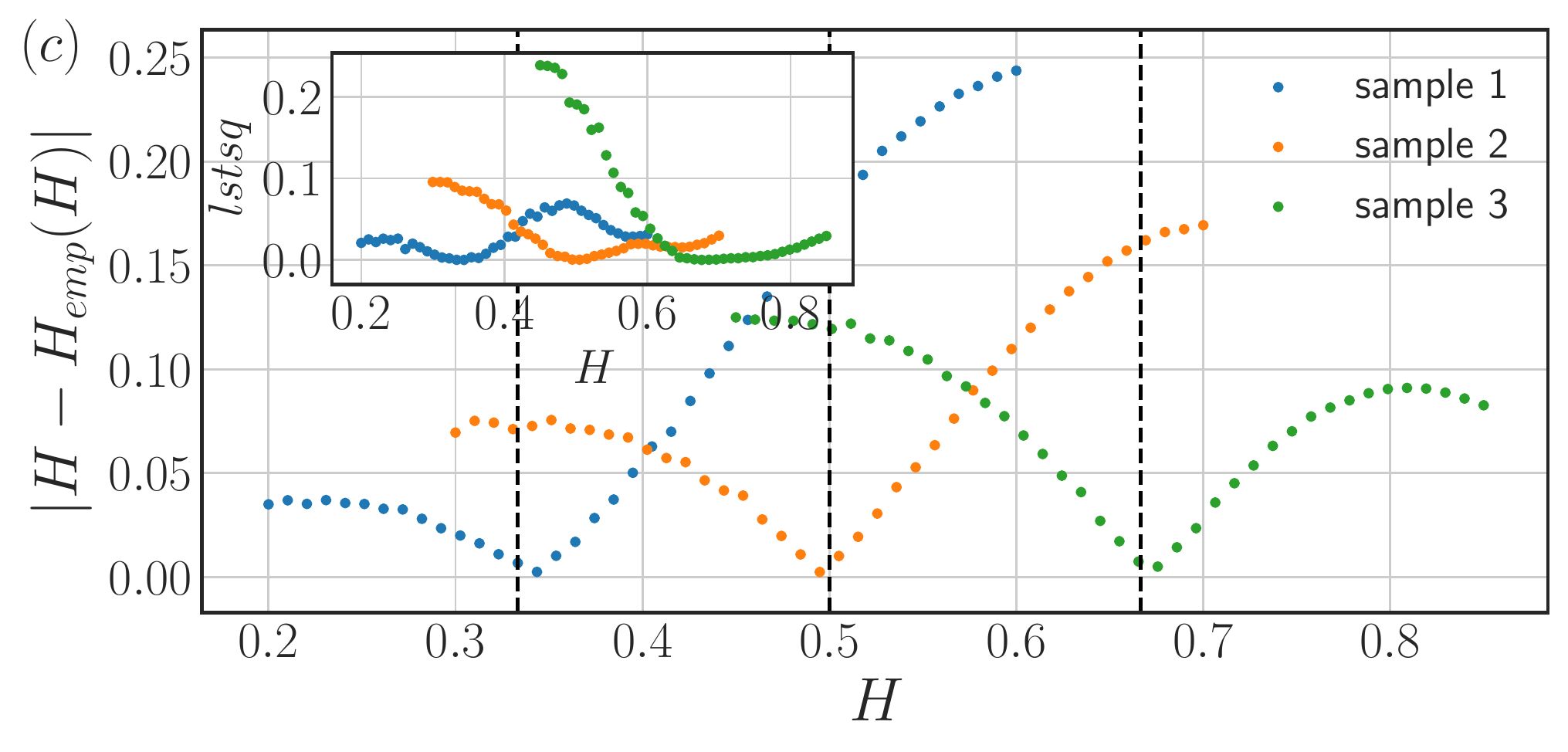}
\caption{(a) Multi-point fractional Brownian bridge (\ref{eq:gen_bridge}) for different Hurst exponents $H$ but identical grid points (black). The $16$ equidistant grid points $(t_i, X_i)$ were drawn randomly from the interval $X_i \in [-0.5,0.5]$. The full temporal resolution comprises $4096$ points
and the inset zooms into the vicinity of one of the prescribed points. (b) Second order structure function $S_2(\tau)= \langle (X^B(t+\tau)-X^B(t))^2 \rangle$ calculated from $100$ realizations of the multi-point fractional Brownian bridge (\ref{eq:gen_bridge}) for $N=32768$ total grid points and $\tilde N=32$ prescribed grid points. Dashed lines correspond to the explicit formula (\ref{eq:s2_exp}), whereas
thin grey lines correspond to the ordinary scaling of fBm $|\tau|^{2H}$ and extend up to the grid length of the coarse grid $\textrm{d}\tilde t=1/32$.
The prescribed grid points were drawn as fBm with $\tilde H=0.5$. Accordingly, the fBb with $H=0.5$ (orange curve) is the \emph{optimal bridge} that belongs to the prescribed points. (c) Cost curve $H_{emp}(H)-H$ for three under-sampled time series $\tilde N=128$ (fBm with $\tilde H=0.33, 0.5, 0.66$ for samples 1-3). Each time series is embedded into bridges (\ref{eq:gen_bridge}) with varying Hurst exponents $H$ and resolution $N=4096$. Therefore, the fit for $H_{emp}$ relies on the left part of (b).}
\label{fig:fBb}
\end{figure}

In the following, we will restrict ourselves to temporal
processes, or more generally to processes depending on only one variable.
The fBm $X(t)$
is a \emph{nonstationary centered Gaussian process}, and thus
entirely characterized by its covariance
\begin{equation}
  \langle X(t)X(t') \rangle = \frac{1}{2}\left(|t|^{2H}+|t'|^{2H}-|t-t'|^{2H}\right)\;.
  \label{eq:fBm_corr}
\end{equation}
which implies that $\langle (X(t) - X(t'))^2 \rangle = |t - t'|^{2H}$, where $H$ denotes the Hurst exponent.
We first present a method to construct a fBm $X(t)$, which
takes specific values $X_i$ at times $t_i$, and subsequently discuss
its application as an optimal stochastic interpolation of a sparsely-sampled real time signal. Accordingly,
we start with the well-known notion of a
fractional Brownian bridge~\cite{delorme:2016,sottinen:2014,Walter:2020aa}, which is defined as fBm starting from zero at $t=0$, ending at $X_1$ at $t=t_1$, \emph{and} possessing the same statistical (including scaling (\ref{eq:fBm_corr})) properties as $X(t)$. Such a fractional Brownian bridge (fBb) can be constructed from $X(t)$ according to
\begin{equation}
  X^B(t)=X(t)- (X(t_1)-X_1) \frac{\langle X(t)X(t_1) \rangle}{\langle X(t_1)^2 \rangle}\;.
  \label{eq:fBb_ordinary}
\end{equation}
It is possible to generalize this ordinary fBb to an arbitrary number of prescribed intermediate grid points in the following manner: First, we consider the $n$-times conditional moments
\begin{align}
  \label{eq:cond1}
  \langle X(t)|\{X_i, t_i\} \rangle =& \frac{\langle X(t) \prod_{i=1}^n \delta(X(t_i)-X_i) \rangle}{\prod_{i=1}^n \langle \delta(X(t_i)-X_i) \rangle}\;,\\
  \langle X(t)X(t')|\{X_i, t_i\}\rangle =& \frac{\langle X(t)X(t') \prod_{i=1}^n \delta(X(t_i)-X_i) \rangle}{\prod_{i=1}^n \langle \delta(X(t_i)-X_i) \rangle}\;.
  \label{eq:cond2}
\end{align}
We then demand that our bridge process $X^B(t)$ is conditional on $X_i$ at $t_i$ for $i=1,\ldots,n$, which is equivalent to the process possessing the conditional moments (\ref{eq:cond1}-\ref{eq:cond2}).
For a Gaussian process with zero mean and covariance
$\langle X(t)X(t') \rangle$, the conditional moments read
%
\begin{equation}
  \langle X(t)|\{X_i, t_i\} \rangle = \left \langle X(t)X(t_i) \right \rangle\sigma_{ij}^{-1}   X_j  \;,
  \label{eq:cond1_calculated}
\end{equation}
and
\begin{align}\label{eq:cond2_calculated}
  \lefteqn{\langle X(t)X(t')|\{X_i, t_i\}\rangle
   = \left \langle X(t)X(t') \right \rangle }\\ \nonumber
  &~~-\left \langle X(t)X(t_i) \right \rangle \left[\sigma_{ij}^{-1} - \sigma_{ik}^{-1} X_k X_l \sigma_{jl}^{-1}  \right]
  \left \langle X(t')X(t_j) \right \rangle\;,
\end{align}
where we implied summation over equal indices and where $\sigma_{ij}=\langle X(t_i)X(t_j)\rangle$ denotes the covariance matrix.
As shown in~\cite{sm},\nocite{lim2002,grebenkov2015multiscale,beran1994statistics,flandrin1989spectrum,fbm,Risken,Mazzolo_2017,stevens1995six,Kolmogorov1962,Oboukhov1962,Friedrich:1997aa,yakhot:2006}
the \emph{multi-point fractional Brownian bridge}
\begin{equation}
  X^B(t) =X(t) -  (X(t_i)-X_i) \sigma_{ij}^{-1}
  \left \langle X(t)X(t_j) \right \rangle\;,
  \label{eq:gen_bridge}
\end{equation}
possesses one- and two-point moments
which are identical to
(\ref{eq:cond1_calculated}-\ref{eq:cond2_calculated}) and we thus conclude that
$X^{B}(t)$ is the stochastic process $X(t)$ conditioned on points $X_i$ at times $t_i$. We indeed obtain $X^B(t_k)= X(t_k)-(X(t_i)-X_i)\sigma_{ij}^{-1}\sigma_{jk}= X(t_k)-(X(t_i)-X_i)\delta_{ik}= X_k$.
Fig.~\ref{fig:fBb}(a) depicts the multi-point fBb (\ref{eq:gen_bridge}) for three different Hurst exponents with $16$ equidistant fixed points (black).
Here, fBm realizations $X(t)$ were generated by the Davies-Harte method~\cite{Davies1987}, which is an exact numerical implementation of fBm (by contrast to other methods, e.g., Fourier representations; see also~\cite{sm} for further specifications).

In order to check that the simulated bridge processes possess the desired properties (\ref{eq:fBm_corr}) and
(\ref{eq:cond1_calculated}-\ref{eq:cond2_calculated}), we have carried out numerical calculations of the second order structure functions
$S_2(\tau)= \left \langle (X^B(t+\tau)-X^B(t))^2 \right \rangle$
for three different Hurst exponents ($H=0.33, 0.5, 0.66$) with a total number of  $N=32768$ total grid points. From these we prescribed $\tilde{N} = 32$ equidistant points generated from fBm (\ref{eq:fBm_corr}) with a Hurst exponent $\tilde{H} = 0.5$.
The results are shown in Fig.~\ref{fig:fBb}(b) and are in agreement with the prediction $S_2(\tau)=|\tau|^{2H}$ (dashed black lines) for small $\tau$. For such a case, where the prescribed points also follow fBm with Hurst exponent $\tilde H$, we can obtain an explicit formula for $S_2(\tau)$ from Eq. (\ref{eq:gen_bridge}), namely
\begin{align}
  \label{eq:s2_exp}
  &S_2(\tau)=
   |\tau|^{2H}\\ \nonumber
   &- \langle \delta_{\tau} X(t) X(t_i)\rangle \left[\sigma_{ij}^{-1}-\sigma_{ik}^{-1}\tilde \sigma_{kl} \sigma_{lj}^{-1}\right]
  \langle
  \delta_{\tau} X(t) X(t_j)\rangle\;,
\end{align}
where $\delta_{\tau} X(t)=X(t+\tau)-X(t)$
and where
$\tilde \sigma_{ij}=\langle X_i X_j\rangle$
denotes the covariance matrix of the prescribed points with $\tilde H$. Consequently, $H=\tilde H$ implies
$\sigma =\tilde \sigma$, which yields $
S_2(\tau)= |\tau|^{2H}$. Hence, Eq. (\ref{eq:s2_exp}) entails that strict self-similarity of the multi-point fBb is only guaranteed if the bridge fBm process $X(t)$ characterized by $H$ matches the Hurst exponent $\tilde H$ of the prescribed points $X_i$. Other bridge multi-point fBbs with $H \neq \tilde H$ correspond to multifractional Brownian motion with a time-dependent Hurst exponent, as shown in Fig.~\ref{fig:fBb}(b)
(we also refer to~\cite{peltier1995multifractional,benassi1997elliptic} for further references).
In other words, given a certain time series $\{X_i,t_i\}$ that possesses a self-similar part governed by $\tilde H$, the bridge with $H=\tilde H$ can be considered as the \emph{optimal stochastic interpolation} of this time series.

Therefore, as already highlighted in the introduction, we are now in the position to describe an optimization procedure that allows estimating Hurst exponents from sparsely sampled time series. The basic idea is to embed a given time series $\{X_i, t_i\}$ into fBbs (\ref{eq:gen_bridge}) with varying Hurst exponents $H$.
For each of these bridges we determine the empirical Hurst exponent $H_{emp}$ as a function of $H$ by fitting the second order structure function up to the smallest time scale of the time series $\textrm{d}\tilde t$,
(i.e., fitting only the left part up to $\textrm{d}t_{coarse}$ in Fig.~\ref{fig:fBb}(b)). This procedure ensures that we only measure deviations from the scaling $|\tau|^{2H}$ in the interpolated region (grey lines in Fig.~\ref{fig:fBb}(b)) and are not directly contaminated by correlations contained in
$\{X_i, t_i\}$.
\begin{figure}[h]
\centering
\includegraphics[width=0.49 \textwidth]{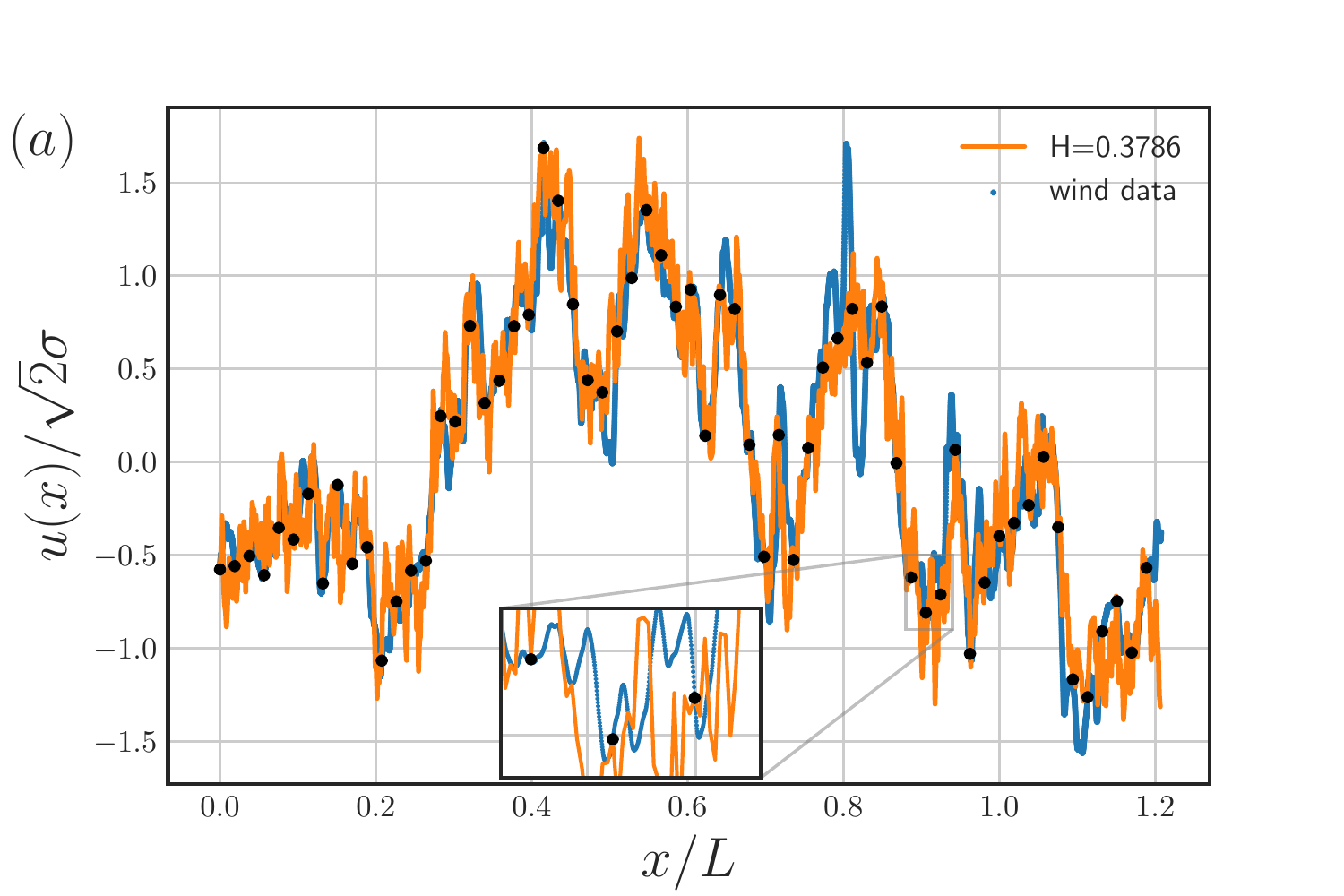}
\includegraphics[width=0.49 \textwidth]{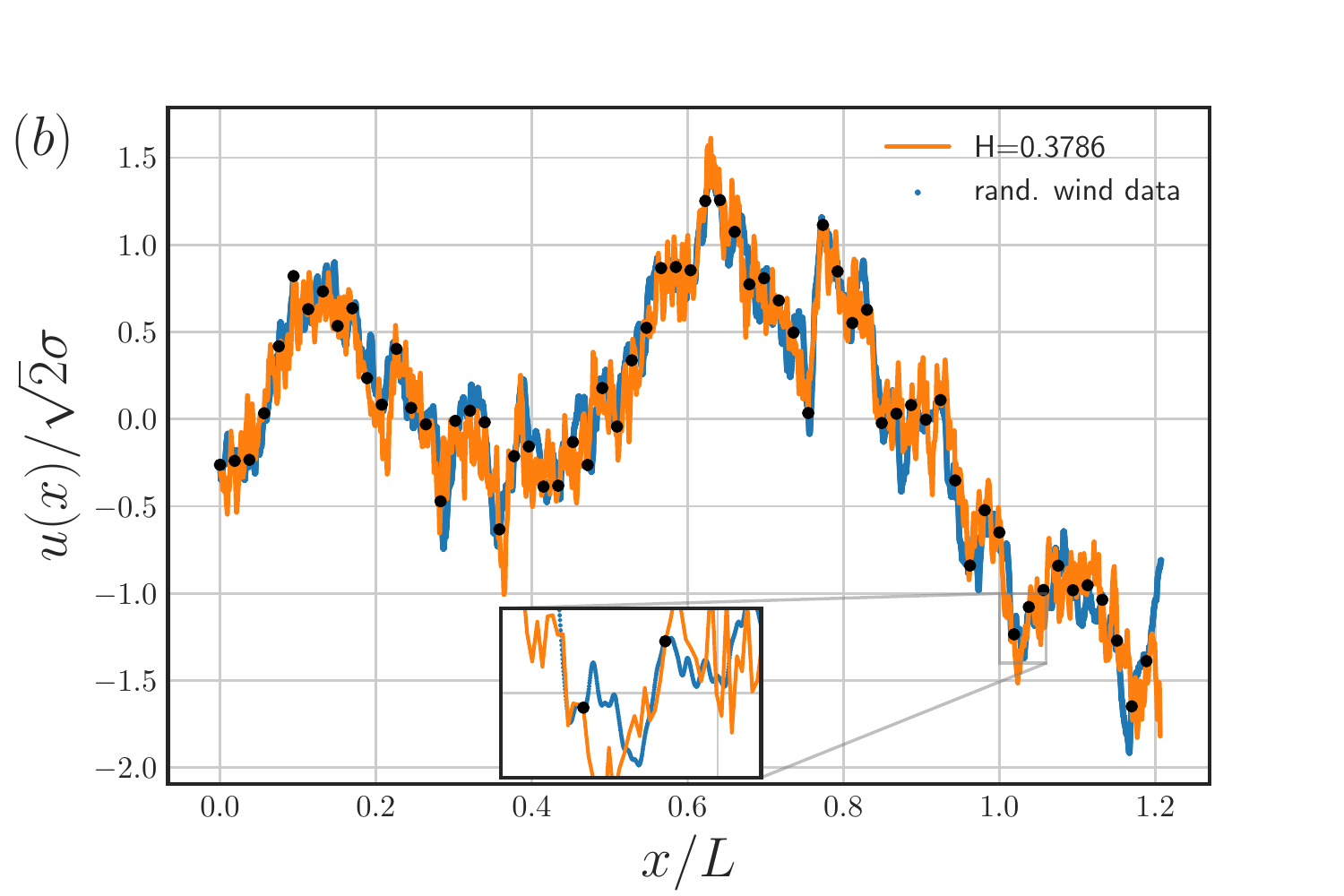}
\caption{(a) Turbulent velocity field measurements (blue) in a von K\'arm\'an experiment using normal Helium. The number of points $N=16384$ corresponds roughly to one integral length scale $L$. The corresponding fBb (orange) was constructed from $\tilde N=64$ points of the signal (black) and possesses a resolution of $1024$ points. The smallest scale of the bridge process therefore corresponds approximately to the Taylor scale of the flow which ensures that the fBb and the velocity field $u(x)$ possess comparable inertial ranges. Small-scale turbulent fluctuations in the velocity field (blue) cannot be reproduced by the fBb due to its restricting Gaussian properties. The Hurst exponent for the fBb $H=0.376$ was determined from randomized samples of the original turbulent signal such as the one depicted in (b). Due to the self-similarity of these samples, an optimization procedure similar to the one depicted in Fig.~\ref{fig:fBb}(c) could be applied. The bridge (orange) is in much better agreement with the self-similar signal (blue) in (b) than with the original signal (blue) in (a).}
\label{fig:wind}
\end{figure}
We have tested our optimization for three different samples of fBm with Hurst exponents $\tilde H=0.33, 0.5, 0.66$ and $\tilde N=128$ grid points. Each of the samples was embedded in fBbs with varying Hurst exponents $H$ and $N=4096$ grid points. Fig.~\ref{fig:fBb}(c) depicts the minimization of $H_{emp}(H)$ for the three samples. It can be clearly seen that optimal Hurst exponents $H_{opt}$ are recovered with high accuracy, although slight variations ($\Delta H_{opt} \approx 0.015$ from 20 different samples of the same $\tilde H$) between different samples can be observed~\cite{sm}. This effect can be attributed to finite sample sizes, and the corresponding deviations remain rather small,
which is quite appealing given the fact that common methods (rescaled range analysis~\cite{Caccia1997} or wavelets~\cite{Simonsen1998}) yield erroneous results for such sparsely sampled time series. The proposed method can thus roughly be considered as the extrapolation of self-similar properties of a given time series to finer scales.

In the examples discussed so far, we have systematically chosen the synthetic signal $X_i$, as well as the process $X(t)$ used in Eqs. (\ref{eq:fBm_corr},\ref{eq:gen_bridge}), to be normalized in the same manner.
In order to apply our optimization to real signals, we examine a turbulent velocity time series
obtained from hot wire anemometry in the superfluid high Reynolds von K\'arm\'an experiment (SHREK) at CEA-Grenoble~\cite{Rousset2014}. The particular experimental setup
is a von K\'arm\'an cell with two-counter rotating disks (-0.12 Hz on top, +0.18 Hz on bottom) in normal Helium (see~\cite{Kharche2018} for further specifications). The temporal resolution is 50kHz and the attained Taylor-Reynolds number was $\textrm{Re}_{\lambda}
=2737$. We applied Taylor's hypothesis of frozen turbulence~\cite{monin} to relate single-point velocity measurement at time $t$ to scales $x=\langle u\rangle t$, where $\langle u \rangle$ is the mean velocity. Furthermore, a key prerequisite for the above mentioned optimization procedure is the standardization of the signal by
$\lim_{r\rightarrow L}
\sqrt{\left \langle (u(x+r)-u(x))^2 \right \rangle}  = \sqrt{2\left \langle u(x))^2 \right \rangle}= \sqrt{2} \sigma$. This standardization ensures the correct large-scale limit of the second-order structure function in Fig.~\ref{fig:fBb}(b), which was necessarily fulfilled by the synthetic samples of fBm (\ref{eq:fBm_corr}) in the previous study.
The blue curve in Fig.~\ref{fig:wind}(a)
depicts an extract of the velocity field $u(x)$ standardized by $\sigma$ of the total signal for around one integral length scale $L$.

By contrast to the above optimization procedure for data from synthetic samples, the present analysis is complicated by: i.) the existence of different scaling regimes in the flow, namely a dissipative and integral range of scales, and ii.) non-self similar (intermittent) features in the signal. Turning to point i.), we chose sample sizes of length $L$ and determined the subset of points in order to guarantee that their grid length lies within the inertial range of scales (here we choose $\approx 100 \eta$).
As far as point ii.) is concerned, intermittent features manifest themselves in form of strongly varying $H_{opt}$ for different samples~\cite{sm}. An example of such an intermittent fluctuation is well visible in the signal shown in Fig.~\ref{fig:wind}(a), at $x \approx 0.8L$. In this region our interpolation procedure does not work very well and the bridge process (\ref{eq:gen_bridge}) has to be generalized to a non-Gaussian process. In this letter, however, we are solely interested in the self-similar part of the signal and thus perform a randomization of Fourier phases of the turbulent signal. A snapshot of the resulting randomized signal is depicted in Fig.~\ref{fig:wind}(b). Strikingly, and contrary to Fig.~\ref{fig:wind}(a), our interpolation procedure leads to much better results than for the original signal. The randomization procedure has effectively suppressed intermittency, and made the signal essentially Gaussian. We fed several samples of this randomized signal into our optimization routine which reduced fluctuations of the Hurst exponent to an extend comparable to the ones observed in synthetic signals~\cite{sm}. Moreover, we obtain a Hurst exponent of the randomized signal
$H_{opt}=0.3786 \pm 0.0251$ (evaluated from the optimization of 100 snapshots~\cite{sm}). To put this result into context, we consider the log-normal model of turbulence~\cite{Kolmogorov1962,Oboukhov1962} which suggests scaling exponents $\zeta_n=n/3-\mu n (n-3)/18$
for the scaling of structure functions $\langle (u(x+r)-u(x))^n \rangle \sim |r|^{\zeta_n}$
where $\mu$ denotes the intermittency coefficient.
Hence, the self-similar part in the K62 model
is given by $H_{K62}= (2+\mu)/6$ and can be compared to our analysis. The above mentioned value from the optimization procedure of the randomized signal $H_{opt}=0.3786 \pm 0.0251$ thus suggests that $\mu = 0.2716 \pm 0.1506$. Latter value for the intermittency coefficient is comparable to $\mu = 0.2913 \pm 0.0853$ acquired from an analysis of the entire turbulent signal~\cite{sm}.
It is thus important to stress here, that the combination of both the Fourier phase randomization and the optimization procedure via the multi-point fBbs is capable of reproducing the monofractal/self-similar part of the turbulent signal, which is also strongly supported by Fig.~\ref{fig:wind}(b).

To conclude, we have presented a generalization of a fractional Brownian bridge to a stochastic process with an arbitrary number of prescribed points.
Furthermore, we devised an optimization method which allowed us to estimate the Hurst exponent of a sparsely sampled time series. Our method
has proven reliable even in the presence of strong anomalous fluctuations, i.e., non-self-similar features, at small scales. In order to address such features, which are visibly not captured by the multi-point fBb in Fig.~\ref{fig:wind}(a), it will be a task for the future to construct multi-point bridge processes with \emph{multiscaling properties} (i.e., which are non-self-similar and potentially possess a dissipative scale~\cite{Chevillard2019,Viggiano2019,rosales2008anomalous,juneja1994synthetic,malara2016fast}).
A generalization of the bridge process (\ref{eq:gen_bridge}) to
an arbitrary number of dimensions is straight-forward and might be of potential interest for the construction of various synthetic fields in several physical contexts.
In turbulence, the full spatio-temporal (though non-intermittent) Eulerian velocity field $\mathbf{u}(\mathbf{x},t)$ can possibly be reconstructed from a set of Lagrangian trajectories $\mathbf{X}(\mathbf{y},t)$ where $\mathbf{\dot X}(\mathbf{y},t)=
\mathbf{u}(\mathbf{X}(\mathbf{y},t),t)$.
Such applications could be of considerable interest for particle tracking measurements, which sometimes require a certain knowledge of the flow field in the vicinity of tracer particles~\cite{Machicoane2019}. Furthermore, the optimization procedure may help to shed light into the ongoing discussion about the inertial range power spectrum in the solar wind~\cite{Boldyrev2005,Goldstein1995}.
To this end, the ideas proposed in this work should be
extended to take into account the existence of several ranges of scales,
involving different power law behaviors~\cite{sm}. Last, we mention possible extensions of our method which might apply or even improve concepts borrowed from machine learning~\cite{brenner2019perspective,doi:10.1190/geo2016-0300.1,APPELHANS201591,rasmussen2003gaussian}. Moreover, the prior assignment of hot- and cold-spot clusters by a simple
$k$-means algorithm might be suitable to model land price fields by multi-point fractional Brownian scalar fields in the widely different domain of urban decision making~\cite{Lengyel2020}.\\
\\
We are grateful to the SHREK collaboration~\cite{Rousset2014,Kharche2018} for providing us with their hot wire anemometry measurements.
J.F. acknowledges funding from the Humboldt Foundation within a Feodor-Lynen fellowship and also benefited from financial support of the Project IDEXLYON of the University of Lyon in the framework of the French program ``Programme Investissements d'Avenir'' (ANR-16-IDEX-0005). S.G.'s visit to Lyon was financially supported by the German Academic Exchange Service through a PROMOS scholarship.

\bibliographystyle{apsrev4-1}
\bibliography{fbm_bridge.bib}

\begin{thebibliography}{72}%
\makeatletter
\providecommand \@ifxundefined [1]{%
 \@ifx{#1\undefined}
}%
\providecommand \@ifnum [1]{%
 \ifnum #1\expandafter \@firstoftwo
 \else \expandafter \@secondoftwo
 \fi
}%
\providecommand \@ifx [1]{%
 \ifx #1\expandafter \@firstoftwo
 \else \expandafter \@secondoftwo
 \fi
}%
\providecommand \natexlab [1]{#1}%
\providecommand \enquote  [1]{``#1''}%
\providecommand \bibnamefont  [1]{#1}%
\providecommand \bibfnamefont [1]{#1}%
\providecommand \citenamefont [1]{#1}%
\providecommand \href@noop [0]{\@secondoftwo}%
\providecommand \href [0]{\begingroup \@sanitize@url \@href}%
\providecommand \@href[1]{\@@startlink{#1}\@@href}%
\providecommand \@@href[1]{\endgroup#1\@@endlink}%
\providecommand \@sanitize@url [0]{\catcode `\\12\catcode `\$12\catcode
  `\&12\catcode `\#12\catcode `\^12\catcode `\_12\catcode `\%12\relax}%
\providecommand \@@startlink[1]{}%
\providecommand \@@endlink[0]{}%
\providecommand \url  [0]{\begingroup\@sanitize@url \@url }%
\providecommand \@url [1]{\endgroup\@href {#1}{\urlprefix }}%
\providecommand \urlprefix  [0]{URL }%
\providecommand \Eprint [0]{\href }%
\providecommand \doibase [0]{http://dx.doi.org/}%
\providecommand \selectlanguage [0]{\@gobble}%
\providecommand \bibinfo  [0]{\@secondoftwo}%
\providecommand \bibfield  [0]{\@secondoftwo}%
\providecommand \translation [1]{[#1]}%
\providecommand \BibitemOpen [0]{}%
\providecommand \bibitemStop [0]{}%
\providecommand \bibitemNoStop [0]{.\EOS\space}%
\providecommand \EOS [0]{\spacefactor3000\relax}%
\providecommand \BibitemShut  [1]{\csname bibitem#1\endcsname}%
\let\auto@bib@innerbib\@empty
\bibitem [{\citenamefont {Haken}(1983)}]{haken1983}%
  \BibitemOpen
  \bibfield  {author} {\bibinfo {author} {\bibfnamefont {H.}~\bibnamefont
  {Haken}},\ }\href@noop {} {\emph {\bibinfo {title} {Synergetics, An
  Introduction}}}\ (\bibinfo  {publisher} {Springer, Berlin, Heidelberg},\
  \bibinfo {year} {1983})\BibitemShut {NoStop}%
\bibitem [{\citenamefont {Prigogine}(1987)}]{prigogineeltit}%
  \BibitemOpen
  \bibfield  {author} {\bibinfo {author} {\bibfnamefont {I.}~\bibnamefont
  {Prigogine}},\ }\href {\doibase https://doi.org/10.1016/0377-2217(87)90085-3}
  {\bibfield  {journal} {\bibinfo  {journal} {Eur. J. Oper. Res.}\ }\textbf
  {\bibinfo {volume} {30}},\ \bibinfo {pages} {97 } (\bibinfo {year}
  {1987})}\BibitemShut {NoStop}%
\bibitem [{\citenamefont {Wax}(1954)}]{Wax}%
  \BibitemOpen
  \bibfield  {author} {\bibinfo {author} {\bibfnamefont {N.}~\bibnamefont
  {Wax}},\ }\href@noop {} {\emph {\bibinfo {title} {Noise and stochastic
  processes}}}\ (\bibinfo  {publisher} {Dover Publications},\ \bibinfo {year}
  {1954})\BibitemShut {NoStop}%
\bibitem [{\citenamefont {Friedrich}\ \emph {et~al.}(2011)\citenamefont
  {Friedrich}, \citenamefont {Peinke}, \citenamefont {Sahimi},\ and\
  \citenamefont {Tabar}}]{Friedrich2011a}%
  \BibitemOpen
  \bibfield  {author} {\bibinfo {author} {\bibfnamefont {R.}~\bibnamefont
  {Friedrich}}, \bibinfo {author} {\bibfnamefont {J.}~\bibnamefont {Peinke}},
  \bibinfo {author} {\bibfnamefont {M.}~\bibnamefont {Sahimi}}, \ and\ \bibinfo
  {author} {\bibfnamefont {R.~M.}\ \bibnamefont {Tabar}},\ }\href {\doibase
  10.1016/j.physrep.2011.05.003} {\bibfield  {journal} {\bibinfo  {journal}
  {Phys. Rep.}\ }\textbf {\bibinfo {volume} {506}},\ \bibinfo {pages} {87}
  (\bibinfo {year} {2011})}\BibitemShut {NoStop}%
\bibitem [{\citenamefont {Einstein}(1905)}]{Einstein1905}%
  \BibitemOpen
  \bibfield  {author} {\bibinfo {author} {\bibfnamefont {A.}~\bibnamefont
  {Einstein}},\ }\href@noop {} {\bibfield  {journal} {\bibinfo  {journal} {Ann.
  Phys.}\ }\textbf {\bibinfo {volume} {17}},\ \bibinfo {pages} {549} (\bibinfo
  {year} {1905})}\BibitemShut {NoStop}%
\bibitem [{\citenamefont {Kolmogorov}(1941)}]{Kolmogorov1941}%
  \BibitemOpen
  \bibfield  {author} {\bibinfo {author} {\bibfnamefont {A.~N.}\ \bibnamefont
  {Kolmogorov}},\ }\href {\doibase 10.1098/rspa.1991.0075} {\bibfield
  {journal} {\bibinfo  {journal} {Dokl. Akad. Nauk Sssr}\ }\textbf {\bibinfo
  {volume} {30}},\ \bibinfo {pages} {301} (\bibinfo {year} {1941})}\BibitemShut
  {NoStop}%
\bibitem [{\citenamefont {Onsager}(1949)}]{Onsager1949}%
  \BibitemOpen
  \bibfield  {author} {\bibinfo {author} {\bibfnamefont {L.}~\bibnamefont
  {Onsager}},\ }\href {\doibase 10.1007/BF02780991} {\bibfield  {journal}
  {\bibinfo  {journal} {Nuovo Cim.}\ }\textbf {\bibinfo {volume} {6}},\
  \bibinfo {pages} {279} (\bibinfo {year} {1949})}\BibitemShut {NoStop}%
\bibitem [{\citenamefont {v.~Weizs{\"{a}}cker}(1948)}]{Weizsacker1948}%
  \BibitemOpen
  \bibfield  {author} {\bibinfo {author} {\bibfnamefont {C.~F.}\ \bibnamefont
  {v.~Weizs{\"{a}}cker}},\ }\href {\doibase 10.1007/BF01668898} {\bibfield
  {journal} {\bibinfo  {journal} {Zeitschrift f{\"{u}}r Phys.}\ }\textbf
  {\bibinfo {volume} {124}},\ \bibinfo {pages} {614} (\bibinfo {year}
  {1948})}\BibitemShut {NoStop}%
\bibitem [{\citenamefont {Goldstein}\ \emph {et~al.}(1995)\citenamefont
  {Goldstein}, \citenamefont {Roberts},\ and\ \citenamefont
  {Matthaeus}}]{Goldstein1995}%
  \BibitemOpen
  \bibfield  {author} {\bibinfo {author} {\bibfnamefont {M.~L.}\ \bibnamefont
  {Goldstein}}, \bibinfo {author} {\bibfnamefont {D.~A.}\ \bibnamefont
  {Roberts}}, \ and\ \bibinfo {author} {\bibfnamefont {W.}~\bibnamefont
  {Matthaeus}},\ }\href@noop {} {\bibfield  {journal} {\bibinfo  {journal}
  {Annu. Rev. Astron. Astrophys.}\ }\textbf {\bibinfo {volume} {33}},\ \bibinfo
  {pages} {283} (\bibinfo {year} {1995})}\BibitemShut {NoStop}%
\bibitem [{\citenamefont {Makarava}\ \emph {et~al.}(2014)\citenamefont
  {Makarava}, \citenamefont {Menz}, \citenamefont {Theves}, \citenamefont
  {Huisinga}, \citenamefont {Beta},\ and\ \citenamefont
  {Holschneider}}]{Makarava2014}%
  \BibitemOpen
  \bibfield  {author} {\bibinfo {author} {\bibfnamefont {N.}~\bibnamefont
  {Makarava}}, \bibinfo {author} {\bibfnamefont {S.}~\bibnamefont {Menz}},
  \bibinfo {author} {\bibfnamefont {M.}~\bibnamefont {Theves}}, \bibinfo
  {author} {\bibfnamefont {W.}~\bibnamefont {Huisinga}}, \bibinfo {author}
  {\bibfnamefont {C.}~\bibnamefont {Beta}}, \ and\ \bibinfo {author}
  {\bibfnamefont {M.}~\bibnamefont {Holschneider}},\ }\href@noop {} {\bibfield
  {journal} {\bibinfo  {journal} {Phys. Rev. E}\ }\textbf {\bibinfo {volume}
  {90}},\ \bibinfo {pages} {042703} (\bibinfo {year} {2014})}\BibitemShut
  {NoStop}%
\bibitem [{\citenamefont {Peng}\ \emph {et~al.}(1993)\citenamefont {Peng},
  \citenamefont {Mietus}, \citenamefont {Hausdorff}, \citenamefont {Havlin},
  \citenamefont {Stanley},\ and\ \citenamefont {Goldberger}}]{Peng1993}%
  \BibitemOpen
  \bibfield  {author} {\bibinfo {author} {\bibfnamefont {C.-K.}\ \bibnamefont
  {Peng}}, \bibinfo {author} {\bibfnamefont {J.}~\bibnamefont {Mietus}},
  \bibinfo {author} {\bibfnamefont {J.~M.}\ \bibnamefont {Hausdorff}}, \bibinfo
  {author} {\bibfnamefont {S.}~\bibnamefont {Havlin}}, \bibinfo {author}
  {\bibfnamefont {H.~E.}\ \bibnamefont {Stanley}}, \ and\ \bibinfo {author}
  {\bibfnamefont {A.~L.}\ \bibnamefont {Goldberger}},\ }\href@noop {}
  {\bibfield  {journal} {\bibinfo  {journal} {Phys. Rev. Lett.}\ }\textbf
  {\bibinfo {volume} {70}},\ \bibinfo {pages} {1343} (\bibinfo {year}
  {1993})}\BibitemShut {NoStop}%
\bibitem [{\citenamefont {L{\'e}vy}\ and\ \citenamefont
  {Loeve}(1965)}]{levy1965processus}%
  \BibitemOpen
  \bibfield  {author} {\bibinfo {author} {\bibfnamefont {P.}~\bibnamefont
  {L{\'e}vy}}\ and\ \bibinfo {author} {\bibfnamefont {M.}~\bibnamefont
  {Loeve}},\ }\href@noop {} {\emph {\bibinfo {title} {Processus stochastiques
  et mouvement brownien}}}\ (\bibinfo  {publisher} {Gauthier-Villars Paris},\
  \bibinfo {year} {1965})\BibitemShut {NoStop}%
\bibitem [{\citenamefont {Mandelbrot}\ and\ \citenamefont
  {Ness}(1968)}]{mandelbrot1968}%
  \BibitemOpen
  \bibfield  {author} {\bibinfo {author} {\bibfnamefont {B.~B.}\ \bibnamefont
  {Mandelbrot}}\ and\ \bibinfo {author} {\bibfnamefont {J.~W.~V.}\ \bibnamefont
  {Ness}},\ }\href {http://www.jstor.org/stable/2027184} {\bibfield  {journal}
  {\bibinfo  {journal} {SIAM Review}\ }\textbf {\bibinfo {volume} {10}},\
  \bibinfo {pages} {422} (\bibinfo {year} {1968})}\BibitemShut {NoStop}%
\bibitem [{\citenamefont {Kranstauber}\ \emph {et~al.}(2012)\citenamefont
  {Kranstauber}, \citenamefont {Kays}, \citenamefont {LaPoint}, \citenamefont
  {Wikelski},\ and\ \citenamefont {Safi}}]{kranstauber2012}%
  \BibitemOpen
  \bibfield  {author} {\bibinfo {author} {\bibfnamefont {B.}~\bibnamefont
  {Kranstauber}}, \bibinfo {author} {\bibfnamefont {R.}~\bibnamefont {Kays}},
  \bibinfo {author} {\bibfnamefont {S.~D.}\ \bibnamefont {LaPoint}}, \bibinfo
  {author} {\bibfnamefont {M.}~\bibnamefont {Wikelski}}, \ and\ \bibinfo
  {author} {\bibfnamefont {K.}~\bibnamefont {Safi}},\ }\href@noop {} {\bibfield
   {journal} {\bibinfo  {journal} {J. Anim. Ecol.}\ }\textbf {\bibinfo {volume}
  {81}},\ \bibinfo {pages} {738} (\bibinfo {year} {2012})}\BibitemShut
  {NoStop}%
\bibitem [{\citenamefont {Brody}\ \emph {et~al.}(2008)\citenamefont {Brody},
  \citenamefont {Hughston},\ and\ \citenamefont {Macrina}}]{Brody2008}%
  \BibitemOpen
  \bibfield  {author} {\bibinfo {author} {\bibfnamefont {D.~C.}\ \bibnamefont
  {Brody}}, \bibinfo {author} {\bibfnamefont {L.~P.}\ \bibnamefont {Hughston}},
  \ and\ \bibinfo {author} {\bibfnamefont {A.}~\bibnamefont {Macrina}},\
  }\href@noop {} {\bibfield  {journal} {\bibinfo  {journal} {Int. J. Theor.
  Appl. Finance}\ }\textbf {\bibinfo {volume} {11}},\ \bibinfo {pages} {107}
  (\bibinfo {year} {2008})}\BibitemShut {NoStop}%
\bibitem [{\citenamefont {de~Mulatier}\ \emph {et~al.}(2015)\citenamefont
  {de~Mulatier}, \citenamefont {Dumonteil}, \citenamefont {Rosso},\ and\
  \citenamefont {Zoia}}]{de_Mulatier_2015}%
  \BibitemOpen
  \bibfield  {author} {\bibinfo {author} {\bibfnamefont {C.}~\bibnamefont
  {de~Mulatier}}, \bibinfo {author} {\bibfnamefont {E.}~\bibnamefont
  {Dumonteil}}, \bibinfo {author} {\bibfnamefont {A.}~\bibnamefont {Rosso}}, \
  and\ \bibinfo {author} {\bibfnamefont {A.}~\bibnamefont {Zoia}},\ }\href
  {\doibase 10.1088/1742-5468/2015/08/p08021} {\bibfield  {journal} {\bibinfo
  {journal} {J. Stat. Mech: Theory Exp.}\ }\textbf {\bibinfo {volume} {2015}},\
  \bibinfo {pages} {P08021} (\bibinfo {year} {2015})}\BibitemShut {NoStop}%
\bibitem [{\citenamefont {Mazzolo}(2017{\natexlab{a}})}]{Mazzolo2017}%
  \BibitemOpen
  \bibfield  {author} {\bibinfo {author} {\bibfnamefont {A.}~\bibnamefont
  {Mazzolo}},\ }\href@noop {} {\bibfield  {journal} {\bibinfo  {journal} {J.
  Stat. Mech: Theory Exp.}\ }\textbf {\bibinfo {volume} {2017}},\ \bibinfo
  {pages} {023203} (\bibinfo {year} {2017}{\natexlab{a}})}\BibitemShut
  {NoStop}%
\bibitem [{\citenamefont {Batista}\ \emph {et~al.}(2016)\citenamefont
  {Batista}, \citenamefont {Dundovic}, \citenamefont {Erdmann}, \citenamefont
  {Kampert}, \citenamefont {Kuempel}, \citenamefont {M{\"u}ller}, \citenamefont
  {Sigl}, \citenamefont {van Vliet}, \citenamefont {Walz},\ and\ \citenamefont
  {Winchen}}]{batista-etal:2016}%
  \BibitemOpen
  \bibfield  {author} {\bibinfo {author} {\bibfnamefont {R.~A.}\ \bibnamefont
  {Batista}}, \bibinfo {author} {\bibfnamefont {A.}~\bibnamefont {Dundovic}},
  \bibinfo {author} {\bibfnamefont {M.}~\bibnamefont {Erdmann}}, \bibinfo
  {author} {\bibfnamefont {K.-H.}\ \bibnamefont {Kampert}}, \bibinfo {author}
  {\bibfnamefont {D.}~\bibnamefont {Kuempel}}, \bibinfo {author} {\bibfnamefont
  {G.}~\bibnamefont {M{\"u}ller}}, \bibinfo {author} {\bibfnamefont
  {G.}~\bibnamefont {Sigl}}, \bibinfo {author} {\bibfnamefont {A.}~\bibnamefont
  {van Vliet}}, \bibinfo {author} {\bibfnamefont {D.}~\bibnamefont {Walz}}, \
  and\ \bibinfo {author} {\bibfnamefont {T.}~\bibnamefont {Winchen}},\ }\href
  {\doibase 10.1088/1475-7516/2016/05/038} {\bibfield  {journal} {\bibinfo
  {journal} {J. Cosmol. Astropart. Phys.}\ }\textbf {\bibinfo {volume}
  {2016}},\ \bibinfo {pages} {038} (\bibinfo {year} {2016})}\BibitemShut
  {NoStop}%
\bibitem [{\citenamefont {Cressie}(1990)}]{cressie1990origins}%
  \BibitemOpen
  \bibfield  {author} {\bibinfo {author} {\bibfnamefont {N.}~\bibnamefont
  {Cressie}},\ }\href@noop {} {\bibfield  {journal} {\bibinfo  {journal}
  {Mathematical geology}\ }\textbf {\bibinfo {volume} {22}},\ \bibinfo {pages}
  {239} (\bibinfo {year} {1990})}\BibitemShut {NoStop}%
\bibitem [{\citenamefont {Delhomme}(1978)}]{Delhomme_1978}%
  \BibitemOpen
  \bibfield  {author} {\bibinfo {author} {\bibfnamefont {J.}~\bibnamefont
  {Delhomme}},\ }\href {\doibase 10.1016/0309-1708(78)90039-8} {\bibfield
  {journal} {\bibinfo  {journal} {Advances in Water Resources}\ }\textbf
  {\bibinfo {volume} {1}},\ \bibinfo {pages} {251} (\bibinfo {year}
  {1978})}\BibitemShut {NoStop}%
\bibitem [{\citenamefont {Gunes}\ and\ \citenamefont
  {Rist}(2008)}]{doi:10.1063/1.3003069}%
  \BibitemOpen
  \bibfield  {author} {\bibinfo {author} {\bibfnamefont {H.}~\bibnamefont
  {Gunes}}\ and\ \bibinfo {author} {\bibfnamefont {U.}~\bibnamefont {Rist}},\
  }\href {\doibase 10.1063/1.3003069} {\bibfield  {journal} {\bibinfo
  {journal} {Phys. Fluids}\ }\textbf {\bibinfo {volume} {20}},\ \bibinfo
  {pages} {104109} (\bibinfo {year} {2008})}\BibitemShut {NoStop}%
\bibitem [{\citenamefont {Hansen}\ and\ \citenamefont
  {Herman}(1989)}]{hansen1989temporal}%
  \BibitemOpen
  \bibfield  {author} {\bibinfo {author} {\bibfnamefont {D.~V.}\ \bibnamefont
  {Hansen}}\ and\ \bibinfo {author} {\bibfnamefont {A.}~\bibnamefont
  {Herman}},\ }\href@noop {} {\bibfield  {journal} {\bibinfo  {journal}
  {Journal of Atmospheric and Oceanic Technology}\ }\textbf {\bibinfo {volume}
  {6}},\ \bibinfo {pages} {599} (\bibinfo {year} {1989})}\BibitemShut {NoStop}%
\bibitem [{\citenamefont {Seo}\ \emph {et~al.}(1990{\natexlab{a}})\citenamefont
  {Seo}, \citenamefont {Krajewski}, \citenamefont {Azimi-Zonooz},\ and\
  \citenamefont {Bowles}}]{seo1990stochastic}%
  \BibitemOpen
  \bibfield  {author} {\bibinfo {author} {\bibfnamefont {D.-J.}\ \bibnamefont
  {Seo}}, \bibinfo {author} {\bibfnamefont {W.~F.}\ \bibnamefont {Krajewski}},
  \bibinfo {author} {\bibfnamefont {A.}~\bibnamefont {Azimi-Zonooz}}, \ and\
  \bibinfo {author} {\bibfnamefont {D.~S.}\ \bibnamefont {Bowles}},\
  }\href@noop {} {\bibfield  {journal} {\bibinfo  {journal} {Water. Resour.
  Res.}\ }\textbf {\bibinfo {volume} {26}},\ \bibinfo {pages} {915} (\bibinfo
  {year} {1990}{\natexlab{a}})}\BibitemShut {NoStop}%
\bibitem [{\citenamefont {Seo}\ \emph {et~al.}(1990{\natexlab{b}})\citenamefont
  {Seo}, \citenamefont {Krajewski},\ and\ \citenamefont
  {Bowles}}]{seo1990stochastic1}%
  \BibitemOpen
  \bibfield  {author} {\bibinfo {author} {\bibfnamefont {D.-J.}\ \bibnamefont
  {Seo}}, \bibinfo {author} {\bibfnamefont {W.~F.}\ \bibnamefont {Krajewski}},
  \ and\ \bibinfo {author} {\bibfnamefont {D.~S.}\ \bibnamefont {Bowles}},\
  }\href@noop {} {\bibfield  {journal} {\bibinfo  {journal} {Water. Resour.
  Res.}\ }\textbf {\bibinfo {volume} {26}},\ \bibinfo {pages} {469} (\bibinfo
  {year} {1990}{\natexlab{b}})}\BibitemShut {NoStop}%
\bibitem [{\citenamefont {Griffa}\ \emph {et~al.}(2004)\citenamefont {Griffa},
  \citenamefont {Piterbarg},\ and\ \citenamefont
  {{\"O}zg{\"o}kmen}}]{griffa2004predictability}%
  \BibitemOpen
  \bibfield  {author} {\bibinfo {author} {\bibfnamefont {A.}~\bibnamefont
  {Griffa}}, \bibinfo {author} {\bibfnamefont {L.~I.}\ \bibnamefont
  {Piterbarg}}, \ and\ \bibinfo {author} {\bibfnamefont {T.}~\bibnamefont
  {{\"O}zg{\"o}kmen}},\ }\href@noop {} {\bibfield  {journal} {\bibinfo
  {journal} {J. Mar. Res.}\ }\textbf {\bibinfo {volume} {62}},\ \bibinfo
  {pages} {1} (\bibinfo {year} {2004})}\BibitemShut {NoStop}%
\bibitem [{\citenamefont {Molz}\ and\ \citenamefont
  {Boman}(1993)}]{molz1993fractal}%
  \BibitemOpen
  \bibfield  {author} {\bibinfo {author} {\bibfnamefont {F.~J.}\ \bibnamefont
  {Molz}}\ and\ \bibinfo {author} {\bibfnamefont {G.~K.}\ \bibnamefont
  {Boman}},\ }\href@noop {} {\bibfield  {journal} {\bibinfo  {journal} {Water.
  Resour. Res.}\ }\textbf {\bibinfo {volume} {29}},\ \bibinfo {pages} {3769}
  (\bibinfo {year} {1993})}\BibitemShut {NoStop}%
\bibitem [{\citenamefont {Molz}\ \emph {et~al.}(1997)\citenamefont {Molz},
  \citenamefont {Liu},\ and\ \citenamefont {Szulga}}]{molz1997fractional}%
  \BibitemOpen
  \bibfield  {author} {\bibinfo {author} {\bibfnamefont {F.}~\bibnamefont
  {Molz}}, \bibinfo {author} {\bibfnamefont {H.}~\bibnamefont {Liu}}, \ and\
  \bibinfo {author} {\bibfnamefont {J.}~\bibnamefont {Szulga}},\ }\href@noop {}
  {\bibfield  {journal} {\bibinfo  {journal} {Water. Resour. Res.}\ }\textbf
  {\bibinfo {volume} {33}},\ \bibinfo {pages} {2273} (\bibinfo {year}
  {1997})}\BibitemShut {NoStop}%
\bibitem [{\citenamefont {Ciesielski}(1961)}]{ciesielski1961holder}%
  \BibitemOpen
  \bibfield  {author} {\bibinfo {author} {\bibfnamefont {Z.}~\bibnamefont
  {Ciesielski}},\ }\href@noop {} {\bibfield  {journal} {\bibinfo  {journal}
  {Trans. Amer. Math. Soc.}\ ,\ \bibinfo {pages} {403}} (\bibinfo {year}
  {1961})}\BibitemShut {NoStop}%
\bibitem [{\citenamefont {Saupe}(1988)}]{saupe1988algorithms}%
  \BibitemOpen
  \bibfield  {author} {\bibinfo {author} {\bibfnamefont {D.}~\bibnamefont
  {Saupe}},\ }in\ \href@noop {} {\emph {\bibinfo {booktitle} {The science of
  fractal images}}}\ (\bibinfo  {publisher} {Springer},\ \bibinfo {year}
  {1988})\ pp.\ \bibinfo {pages} {71--136}\BibitemShut {NoStop}%
\bibitem [{\citenamefont {Voss}(1988)}]{Voss_1988}%
  \BibitemOpen
  \bibfield  {author} {\bibinfo {author} {\bibfnamefont {R.~F.}\ \bibnamefont
  {Voss}},\ }\href {\doibase 10.1007/978-1-4612-3784-6_1} {\bibfield  {journal}
  {\bibinfo  {journal} {The Science of Fractal Images}\ ,\ \bibinfo {pages}
  {21}} (\bibinfo {year} {1988})}\BibitemShut {NoStop}%
\bibitem [{\citenamefont {Schlickeiser}(2015)}]{schlickeiser:2015}%
  \BibitemOpen
  \bibfield  {author} {\bibinfo {author} {\bibfnamefont {R.}~\bibnamefont
  {Schlickeiser}},\ }\href {\doibase 10.1063/1.4928940} {\bibfield  {journal}
  {\bibinfo  {journal} {Phys. Plasmas}\ }\textbf {\bibinfo {volume} {22}},\
  \bibinfo {pages} {091502} (\bibinfo {year} {2015})}\BibitemShut {NoStop}%
\bibitem [{\citenamefont {Zweibel}(2013)}]{zweibel:2013}%
  \BibitemOpen
  \bibfield  {author} {\bibinfo {author} {\bibfnamefont {E.~G.}\ \bibnamefont
  {Zweibel}},\ }\href {\doibase 10.1063/1.4807033} {\bibfield  {journal}
  {\bibinfo  {journal} {Phys. Plasmas}\ }\textbf {\bibinfo {volume} {20}},\
  \bibinfo {pages} {055501} (\bibinfo {year} {2013})}\BibitemShut {NoStop}%
\bibitem [{\citenamefont {Giacalone}\ and\ \citenamefont
  {Jokipii}(1999)}]{giacalone-jokipii:1999}%
  \BibitemOpen
  \bibfield  {author} {\bibinfo {author} {\bibfnamefont {J.}~\bibnamefont
  {Giacalone}}\ and\ \bibinfo {author} {\bibfnamefont {J.~R.}\ \bibnamefont
  {Jokipii}},\ }\href {http://stacks.iop.org/0004-637X/520/i=1/a=204}
  {\bibfield  {journal} {\bibinfo  {journal} {ApJ}\ }\textbf {\bibinfo {volume}
  {520}},\ \bibinfo {pages} {204} (\bibinfo {year} {1999})}\BibitemShut
  {NoStop}%
\bibitem [{\citenamefont {{Zimbardo}}\ \emph {et~al.}(2015)\citenamefont
  {{Zimbardo}}, \citenamefont {{Amato}}, \citenamefont {{Bovet}}, \citenamefont
  {{Effenberger}}, \citenamefont {{Fasoli}}, \citenamefont {{Fichtner}},
  \citenamefont {{Furno}}, \citenamefont {{Gustafson}}, \citenamefont
  {{Ricci}},\ and\ \citenamefont {{Perri}}}]{zimbardo-amato-etal:2015}%
  \BibitemOpen
  \bibfield  {author} {\bibinfo {author} {\bibfnamefont {G.}~\bibnamefont
  {{Zimbardo}}}, \bibinfo {author} {\bibfnamefont {E.}~\bibnamefont {{Amato}}},
  \bibinfo {author} {\bibfnamefont {A.}~\bibnamefont {{Bovet}}}, \bibinfo
  {author} {\bibfnamefont {F.}~\bibnamefont {{Effenberger}}}, \bibinfo {author}
  {\bibfnamefont {A.}~\bibnamefont {{Fasoli}}}, \bibinfo {author}
  {\bibfnamefont {H.}~\bibnamefont {{Fichtner}}}, \bibinfo {author}
  {\bibfnamefont {I.}~\bibnamefont {{Furno}}}, \bibinfo {author} {\bibfnamefont
  {K.}~\bibnamefont {{Gustafson}}}, \bibinfo {author} {\bibfnamefont
  {P.}~\bibnamefont {{Ricci}}}, \ and\ \bibinfo {author} {\bibfnamefont
  {S.}~\bibnamefont {{Perri}}},\ }\href {\doibase 10.1017/S0022377815001117}
  {\bibfield  {journal} {\bibinfo  {journal} {J. Plasma Phys.}\ }\textbf
  {\bibinfo {volume} {81}},\ \bibinfo {eid} {495810601} (\bibinfo {year}
  {2015})}\BibitemShut {NoStop}%
\bibitem [{\citenamefont {Snodin}\ \emph {et~al.}(2016)\citenamefont {Snodin},
  \citenamefont {Shukurov}, \citenamefont {Sarson}, \citenamefont {Bushby},\
  and\ \citenamefont {Rodrigues}}]{snodin-shukurov-etal:2016}%
  \BibitemOpen
  \bibfield  {author} {\bibinfo {author} {\bibfnamefont {A.~P.}\ \bibnamefont
  {Snodin}}, \bibinfo {author} {\bibfnamefont {A.}~\bibnamefont {Shukurov}},
  \bibinfo {author} {\bibfnamefont {G.~R.}\ \bibnamefont {Sarson}}, \bibinfo
  {author} {\bibfnamefont {P.~J.}\ \bibnamefont {Bushby}}, \ and\ \bibinfo
  {author} {\bibfnamefont {L.~F.~S.}\ \bibnamefont {Rodrigues}},\ }\href@noop
  {} {\bibfield  {journal} {\bibinfo  {journal} {MNRAS}\ }\textbf {\bibinfo
  {volume} {457}},\ \bibinfo {pages} {3975} (\bibinfo {year}
  {2016})}\BibitemShut {NoStop}%
\bibitem [{\citenamefont {{Reichherzer}}\ \emph {et~al.}(2019)\citenamefont
  {{Reichherzer}}, \citenamefont {{Becker Tjus}}, \citenamefont {{Zweibel}},
  \citenamefont {{Merten}},\ and\ \citenamefont
  {{Pueschel}}}]{reichherzer-etal:2019}%
  \BibitemOpen
  \bibfield  {author} {\bibinfo {author} {\bibfnamefont {P.}~\bibnamefont
  {{Reichherzer}}}, \bibinfo {author} {\bibfnamefont {J.}~\bibnamefont {{Becker
  Tjus}}}, \bibinfo {author} {\bibfnamefont {E.~G.}\ \bibnamefont {{Zweibel}}},
  \bibinfo {author} {\bibfnamefont {L.}~\bibnamefont {{Merten}}}, \ and\
  \bibinfo {author} {\bibfnamefont {M.~J.}\ \bibnamefont {{Pueschel}}},\
  }\href@noop {} {\bibfield  {journal} {\bibinfo  {journal} {arXiv e-prints}\
  ,\ \bibinfo {eid} {arXiv:1910.07528}} (\bibinfo {year} {2019})}\BibitemShut
  {NoStop}%
\bibitem [{\citenamefont {Delorme}\ and\ \citenamefont
  {Wiese}(2016)}]{delorme:2016}%
  \BibitemOpen
  \bibfield  {author} {\bibinfo {author} {\bibfnamefont {M.}~\bibnamefont
  {Delorme}}\ and\ \bibinfo {author} {\bibfnamefont {K.~J.}\ \bibnamefont
  {Wiese}},\ }\href@noop {} {\bibfield  {journal} {\bibinfo  {journal} {Phys.
  Rev. E}\ }\textbf {\bibinfo {volume} {94}},\ \bibinfo {pages} {052105}
  (\bibinfo {year} {2016})}\BibitemShut {NoStop}%
\bibitem [{\citenamefont {Sottinen}\ and\ \citenamefont
  {Yazigi}(2014)}]{sottinen:2014}%
  \BibitemOpen
  \bibfield  {author} {\bibinfo {author} {\bibfnamefont {T.}~\bibnamefont
  {Sottinen}}\ and\ \bibinfo {author} {\bibfnamefont {A.}~\bibnamefont
  {Yazigi}},\ }\href {\doibase https://doi.org/10.1016/j.spa.2014.04.002}
  {\bibfield  {journal} {\bibinfo  {journal} {Stochastic Processes Appl.}\
  }\textbf {\bibinfo {volume} {124}},\ \bibinfo {pages} {3084 } (\bibinfo
  {year} {2014})}\BibitemShut {NoStop}%
\bibitem [{\citenamefont {Walter}\ and\ \citenamefont
  {Wiese}(2020)}]{Walter:2020aa}%
  \BibitemOpen
  \bibfield  {author} {\bibinfo {author} {\bibfnamefont {B.}~\bibnamefont
  {Walter}}\ and\ \bibinfo {author} {\bibfnamefont {K.~J.}\ \bibnamefont
  {Wiese}},\ }\href@noop {} {\bibfield  {journal} {\bibinfo  {journal} {Phys.
  Rev. E}\ }\textbf {\bibinfo {volume} {101}} (\bibinfo {year}
  {2020})}\BibitemShut {NoStop}%
\bibitem [{sm()}]{sm}%
  \BibitemOpen
  \href@noop {} {}\bibinfo {howpublished} {See Supplemental Material, which
  includes Refs. [41-52], for further discussion and proofs.}\BibitemShut
  {Stop}%
\bibitem [{\citenamefont {Lim}\ and\ \citenamefont {Muniandy}(2002)}]{lim2002}%
  \BibitemOpen
  \bibfield  {author} {\bibinfo {author} {\bibfnamefont {S.~C.}\ \bibnamefont
  {Lim}}\ and\ \bibinfo {author} {\bibfnamefont {S.~V.}\ \bibnamefont
  {Muniandy}},\ }\href {\doibase 10.1103/PhysRevE.66.021114} {\bibfield
  {journal} {\bibinfo  {journal} {Phys. Rev. E}\ }\textbf {\bibinfo {volume}
  {66}},\ \bibinfo {pages} {021114} (\bibinfo {year} {2002})}\BibitemShut
  {NoStop}%
\bibitem [{\citenamefont {Grebenkov}\ \emph {et~al.}(2015)\citenamefont
  {Grebenkov}, \citenamefont {Belyaev},\ and\ \citenamefont
  {Jones}}]{grebenkov2015multiscale}%
  \BibitemOpen
  \bibfield  {author} {\bibinfo {author} {\bibfnamefont {D.~S.}\ \bibnamefont
  {Grebenkov}}, \bibinfo {author} {\bibfnamefont {D.}~\bibnamefont {Belyaev}},
  \ and\ \bibinfo {author} {\bibfnamefont {P.~W.}\ \bibnamefont {Jones}},\
  }\href@noop {} {\bibfield  {journal} {\bibinfo  {journal} {Journal of Physics
  A: Mathematical and Theoretical}\ }\textbf {\bibinfo {volume} {49}},\
  \bibinfo {pages} {043001} (\bibinfo {year} {2015})}\BibitemShut {NoStop}%
\bibitem [{\citenamefont {Beran}(1994)}]{beran1994statistics}%
  \BibitemOpen
  \bibfield  {author} {\bibinfo {author} {\bibfnamefont {J.}~\bibnamefont
  {Beran}},\ }\href@noop {} {\emph {\bibinfo {title} {Statistics for
  long-memory processes}}},\ Vol.~\bibinfo {volume} {61}\ (\bibinfo
  {publisher} {CRC press},\ \bibinfo {year} {1994})\BibitemShut {NoStop}%
\bibitem [{\citenamefont {Flandrin}(1989)}]{flandrin1989spectrum}%
  \BibitemOpen
  \bibfield  {author} {\bibinfo {author} {\bibfnamefont {P.}~\bibnamefont
  {Flandrin}},\ }\href@noop {} {\bibfield  {journal} {\bibinfo  {journal} {IEEE
  Transactions on information theory}\ }\textbf {\bibinfo {volume} {35}},\
  \bibinfo {pages} {197} (\bibinfo {year} {1989})}\BibitemShut {NoStop}%
\bibitem [{fbm()}]{fbm}%
  \BibitemOpen
  \href@noop {} {}\bibinfo {howpublished}
  {https://pypi.org/project/fbm/}\BibitemShut {NoStop}%
\bibitem [{\citenamefont {Risken}(1996)}]{Risken}%
  \BibitemOpen
  \bibfield  {author} {\bibinfo {author} {\bibfnamefont {H.}~\bibnamefont
  {Risken}},\ }\href@noop {} {\emph {\bibinfo {title} {{The Fokker-Planck
  Equation}}}}\ (\bibinfo  {publisher} {Springer, Berlin},\ \bibinfo {year}
  {1996})\BibitemShut {NoStop}%
\bibitem [{\citenamefont {Mazzolo}(2017{\natexlab{b}})}]{Mazzolo_2017}%
  \BibitemOpen
  \bibfield  {author} {\bibinfo {author} {\bibfnamefont {A.}~\bibnamefont
  {Mazzolo}},\ }\href {\doibase 10.1063/1.5000077} {\bibfield  {journal}
  {\bibinfo  {journal} {J. Math. Phys.}\ }\textbf {\bibinfo {volume} {58}},\
  \bibinfo {pages} {093302} (\bibinfo {year} {2017}{\natexlab{b}})}\BibitemShut
  {NoStop}%
\bibitem [{\citenamefont {Stevens}(1995)}]{stevens1995six}%
  \BibitemOpen
  \bibfield  {author} {\bibinfo {author} {\bibfnamefont {C.~F.}\ \bibnamefont
  {Stevens}},\ }\href@noop {} {\emph {\bibinfo {title} {The six core theories
  of modern physics}}}\ (\bibinfo  {publisher} {MIT Press},\ \bibinfo {year}
  {1995})\BibitemShut {NoStop}%
\bibitem [{\citenamefont {Kolmogorov}(1962)}]{Kolmogorov1962}%
  \BibitemOpen
  \bibfield  {author} {\bibinfo {author} {\bibfnamefont {A.~N.}\ \bibnamefont
  {Kolmogorov}},\ }\href {\doibase 10.1017/S0022112062000518} {\bibfield
  {journal} {\bibinfo  {journal} {J. Fluid Mech.}\ }\textbf {\bibinfo {volume}
  {13}},\ \bibinfo {pages} {82} (\bibinfo {year} {1962})}\BibitemShut {NoStop}%
\bibitem [{\citenamefont {Oboukhov}(1962)}]{Oboukhov1962}%
  \BibitemOpen
  \bibfield  {author} {\bibinfo {author} {\bibfnamefont {A.~M.}\ \bibnamefont
  {Oboukhov}},\ }\href {\doibase 10.1017/S0022112062000506} {\bibfield
  {journal} {\bibinfo  {journal} {J. Fluid Mech.}\ }\textbf {\bibinfo {volume}
  {67}},\ \bibinfo {pages} {77} (\bibinfo {year} {1962})}\BibitemShut {NoStop}%
\bibitem [{\citenamefont {Friedrich}\ and\ \citenamefont
  {Peinke}(1997)}]{Friedrich:1997aa}%
  \BibitemOpen
  \bibfield  {author} {\bibinfo {author} {\bibfnamefont {R.}~\bibnamefont
  {Friedrich}}\ and\ \bibinfo {author} {\bibfnamefont {J.}~\bibnamefont
  {Peinke}},\ }\href {\doibase 10.1103/PhysRevLett.78.863} {\bibfield
  {journal} {\bibinfo  {journal} {Phys. Rev. Lett.}\ }\textbf {\bibinfo
  {volume} {78}},\ \bibinfo {pages} {863} (\bibinfo {year} {1997})}\BibitemShut
  {NoStop}%
\bibitem [{\citenamefont {Yakhot}(2006)}]{yakhot:2006}%
  \BibitemOpen
  \bibfield  {author} {\bibinfo {author} {\bibfnamefont {V.}~\bibnamefont
  {Yakhot}},\ }\href@noop {} {\bibfield  {journal} {\bibinfo  {journal} {Phys.
  D}\ }\textbf {\bibinfo {volume} {215}},\ \bibinfo {pages} {166} (\bibinfo
  {year} {2006})}\BibitemShut {NoStop}%
\bibitem [{\citenamefont {Davies}\ and\ \citenamefont
  {Harte}(1987)}]{Davies1987}%
  \BibitemOpen
  \bibfield  {author} {\bibinfo {author} {\bibfnamefont {R.~B.}\ \bibnamefont
  {Davies}}\ and\ \bibinfo {author} {\bibfnamefont {D.~S.}\ \bibnamefont
  {Harte}},\ }\href@noop {} {\bibfield  {journal} {\bibinfo  {journal}
  {Biometrika}\ }\textbf {\bibinfo {volume} {74}},\ \bibinfo {pages} {95}
  (\bibinfo {year} {1987})}\BibitemShut {NoStop}%
\bibitem [{\citenamefont {Peltier}\ and\ \citenamefont
  {V{\'e}hel}(1995)}]{peltier1995multifractional}%
  \BibitemOpen
  \bibfield  {author} {\bibinfo {author} {\bibfnamefont {R.-F.}\ \bibnamefont
  {Peltier}}\ and\ \bibinfo {author} {\bibfnamefont {J.~L.}\ \bibnamefont
  {V{\'e}hel}},\ }\href@noop {} {\bibfield  {journal} {\bibinfo  {journal}
  {INRIA, Project 2645}\ } (\bibinfo {year} {1995})}\BibitemShut {NoStop}%
\bibitem [{\citenamefont {Benassi}\ \emph {et~al.}(1997)\citenamefont
  {Benassi}, \citenamefont {Roux},\ and\ \citenamefont
  {Jaffard}}]{benassi1997elliptic}%
  \BibitemOpen
  \bibfield  {author} {\bibinfo {author} {\bibfnamefont {A.}~\bibnamefont
  {Benassi}}, \bibinfo {author} {\bibfnamefont {D.}~\bibnamefont {Roux}}, \
  and\ \bibinfo {author} {\bibfnamefont {S.}~\bibnamefont {Jaffard}},\
  }\href@noop {} {\bibfield  {journal} {\bibinfo  {journal} {Rev. Mat.
  Iberoam.}\ }\textbf {\bibinfo {volume} {13}},\ \bibinfo {pages} {19}
  (\bibinfo {year} {1997})}\BibitemShut {NoStop}%
\bibitem [{\citenamefont {Caccia}\ \emph {et~al.}(1997)\citenamefont {Caccia},
  \citenamefont {Percival}, \citenamefont {Cannon}, \citenamefont {Raymond},\
  and\ \citenamefont {Bassingthwaighte}}]{Caccia1997}%
  \BibitemOpen
  \bibfield  {author} {\bibinfo {author} {\bibfnamefont {D.~C.}\ \bibnamefont
  {Caccia}}, \bibinfo {author} {\bibfnamefont {D.}~\bibnamefont {Percival}},
  \bibinfo {author} {\bibfnamefont {M.~J.}\ \bibnamefont {Cannon}}, \bibinfo
  {author} {\bibfnamefont {G.}~\bibnamefont {Raymond}}, \ and\ \bibinfo
  {author} {\bibfnamefont {J.~B.}\ \bibnamefont {Bassingthwaighte}},\
  }\href@noop {} {\bibfield  {journal} {\bibinfo  {journal} {Physica A}\
  }\textbf {\bibinfo {volume} {246}},\ \bibinfo {pages} {609} (\bibinfo {year}
  {1997})}\BibitemShut {NoStop}%
\bibitem [{\citenamefont {Simonsen}\ \emph {et~al.}(1998)\citenamefont
  {Simonsen}, \citenamefont {Hansen},\ and\ \citenamefont
  {Nes}}]{Simonsen1998}%
  \BibitemOpen
  \bibfield  {author} {\bibinfo {author} {\bibfnamefont {I.}~\bibnamefont
  {Simonsen}}, \bibinfo {author} {\bibfnamefont {A.}~\bibnamefont {Hansen}}, \
  and\ \bibinfo {author} {\bibfnamefont {O.~M.}\ \bibnamefont {Nes}},\
  }\href@noop {} {\bibfield  {journal} {\bibinfo  {journal} {Phys. Rev. E}\
  }\textbf {\bibinfo {volume} {58}},\ \bibinfo {pages} {2779} (\bibinfo {year}
  {1998})}\BibitemShut {NoStop}%
\bibitem [{\citenamefont {Rousset}\ \emph {et~al.}(2014)\citenamefont
  {Rousset}, \citenamefont {Bonnay}, \citenamefont {Diribarne}, \citenamefont
  {Girard}, \citenamefont {Poncet}, \citenamefont {Herbert}, \citenamefont
  {Salort}, \citenamefont {Baudet}, \citenamefont {Castaing}, \citenamefont
  {Chevillard} \emph {et~al.}}]{Rousset2014}%
  \BibitemOpen
  \bibfield  {author} {\bibinfo {author} {\bibfnamefont {B.}~\bibnamefont
  {Rousset}}, \bibinfo {author} {\bibfnamefont {P.}~\bibnamefont {Bonnay}},
  \bibinfo {author} {\bibfnamefont {P.}~\bibnamefont {Diribarne}}, \bibinfo
  {author} {\bibfnamefont {A.}~\bibnamefont {Girard}}, \bibinfo {author}
  {\bibfnamefont {J.-M.}\ \bibnamefont {Poncet}}, \bibinfo {author}
  {\bibfnamefont {E.}~\bibnamefont {Herbert}}, \bibinfo {author} {\bibfnamefont
  {J.}~\bibnamefont {Salort}}, \bibinfo {author} {\bibfnamefont
  {C.}~\bibnamefont {Baudet}}, \bibinfo {author} {\bibfnamefont
  {B.}~\bibnamefont {Castaing}}, \bibinfo {author} {\bibfnamefont
  {L.}~\bibnamefont {Chevillard}},  \emph {et~al.},\ }\href@noop {} {\bibfield
  {journal} {\bibinfo  {journal} {Rev. Sci. Instrum.}\ }\textbf {\bibinfo
  {volume} {85}},\ \bibinfo {pages} {103908} (\bibinfo {year}
  {2014})}\BibitemShut {NoStop}%
\bibitem [{\citenamefont {Kharche}\ \emph {et~al.}(2018)\citenamefont
  {Kharche}, \citenamefont {Bon-Mardion}, \citenamefont {Moro}, \citenamefont
  {Peinke}, \citenamefont {Rousset},\ and\ \citenamefont
  {Girard}}]{Kharche2018}%
  \BibitemOpen
  \bibfield  {author} {\bibinfo {author} {\bibfnamefont {S.}~\bibnamefont
  {Kharche}}, \bibinfo {author} {\bibfnamefont {M.}~\bibnamefont
  {Bon-Mardion}}, \bibinfo {author} {\bibfnamefont {J.-P.}\ \bibnamefont
  {Moro}}, \bibinfo {author} {\bibfnamefont {J.}~\bibnamefont {Peinke}},
  \bibinfo {author} {\bibfnamefont {B.}~\bibnamefont {Rousset}}, \ and\
  \bibinfo {author} {\bibfnamefont {A.}~\bibnamefont {Girard}},\ }in\
  \href@noop {} {\emph {\bibinfo {booktitle} {iTi Conference on Turbulence}}}\
  (\bibinfo {organization} {Springer},\ \bibinfo {year} {2018})\ pp.\ \bibinfo
  {pages} {179--184}\BibitemShut {NoStop}%
\bibitem [{\citenamefont {Monin}\ and\ \citenamefont {Yaglom}(2007)}]{monin}%
  \BibitemOpen
  \bibfield  {author} {\bibinfo {author} {\bibfnamefont {A.~S.}\ \bibnamefont
  {Monin}}\ and\ \bibinfo {author} {\bibfnamefont {A.~M.}\ \bibnamefont
  {Yaglom}},\ }\href@noop {} {\emph {\bibinfo {title} {{Statistical Fluid
  Mechanics: Mechanics of Turbulence}}}}\ (\bibinfo  {publisher} {Courier Dover
  Publications},\ \bibinfo {year} {2007})\BibitemShut {NoStop}%
\bibitem [{\citenamefont {Chevillard}\ \emph {et~al.}(2019)\citenamefont
  {Chevillard}, \citenamefont {Garban}, \citenamefont {Rhodes},\ and\
  \citenamefont {Vargas}}]{Chevillard2019}%
  \BibitemOpen
  \bibfield  {author} {\bibinfo {author} {\bibfnamefont {L.}~\bibnamefont
  {Chevillard}}, \bibinfo {author} {\bibfnamefont {C.}~\bibnamefont {Garban}},
  \bibinfo {author} {\bibfnamefont {R.}~\bibnamefont {Rhodes}}, \ and\ \bibinfo
  {author} {\bibfnamefont {V.}~\bibnamefont {Vargas}},\ }in\ \href@noop {}
  {\emph {\bibinfo {booktitle} {Annales Henri Poincar{\'e}}}},\ Vol.~\bibinfo
  {volume} {20}\ (\bibinfo {organization} {Springer},\ \bibinfo {year} {2019})\
  pp.\ \bibinfo {pages} {3693--3741}\BibitemShut {NoStop}%
\bibitem [{\citenamefont {Viggiano}\ \emph {et~al.}(2020)\citenamefont
  {Viggiano}, \citenamefont {Friedrich}, \citenamefont {Volk}, \citenamefont
  {Bourgoin}, \citenamefont {Cal},\ and\ \citenamefont
  {Chevillard}}]{Viggiano2019}%
  \BibitemOpen
  \bibfield  {author} {\bibinfo {author} {\bibfnamefont {B.}~\bibnamefont
  {Viggiano}}, \bibinfo {author} {\bibfnamefont {J.}~\bibnamefont {Friedrich}},
  \bibinfo {author} {\bibfnamefont {R.}~\bibnamefont {Volk}}, \bibinfo {author}
  {\bibfnamefont {M.}~\bibnamefont {Bourgoin}}, \bibinfo {author}
  {\bibfnamefont {R.~B.}\ \bibnamefont {Cal}}, \ and\ \bibinfo {author}
  {\bibfnamefont {L.}~\bibnamefont {Chevillard}},\ }\href
  {http://dx.doi.org/10.1017/jfm.2020.495} {\bibfield  {journal} {\bibinfo
  {journal} {J. Fluid Mech.}\ }\textbf {\bibinfo {volume} {900}} (\bibinfo
  {year} {2020})}\BibitemShut {NoStop}%
\bibitem [{\citenamefont {Rosales}\ and\ \citenamefont
  {Meneveau}(2008)}]{rosales2008anomalous}%
  \BibitemOpen
  \bibfield  {author} {\bibinfo {author} {\bibfnamefont {C.}~\bibnamefont
  {Rosales}}\ and\ \bibinfo {author} {\bibfnamefont {C.}~\bibnamefont
  {Meneveau}},\ }\href@noop {} {\bibfield  {journal} {\bibinfo  {journal}
  {Phys. Rev. E}\ }\textbf {\bibinfo {volume} {78}},\ \bibinfo {pages} {016313}
  (\bibinfo {year} {2008})}\BibitemShut {NoStop}%
\bibitem [{\citenamefont {Juneja}\ \emph {et~al.}(1994)\citenamefont {Juneja},
  \citenamefont {Lathrop}, \citenamefont {Sreenivasan},\ and\ \citenamefont
  {Stolovitzky}}]{juneja1994synthetic}%
  \BibitemOpen
  \bibfield  {author} {\bibinfo {author} {\bibfnamefont {A.}~\bibnamefont
  {Juneja}}, \bibinfo {author} {\bibfnamefont {D.}~\bibnamefont {Lathrop}},
  \bibinfo {author} {\bibfnamefont {K.}~\bibnamefont {Sreenivasan}}, \ and\
  \bibinfo {author} {\bibfnamefont {G.}~\bibnamefont {Stolovitzky}},\
  }\href@noop {} {\bibfield  {journal} {\bibinfo  {journal} {Phys. Rev. E}\
  }\textbf {\bibinfo {volume} {49}},\ \bibinfo {pages} {5179} (\bibinfo {year}
  {1994})}\BibitemShut {NoStop}%
\bibitem [{\citenamefont {Malara}\ \emph {et~al.}(2016)\citenamefont {Malara},
  \citenamefont {Di~Mare}, \citenamefont {Nigro},\ and\ \citenamefont
  {Sorriso-Valvo}}]{malara2016fast}%
  \BibitemOpen
  \bibfield  {author} {\bibinfo {author} {\bibfnamefont {F.}~\bibnamefont
  {Malara}}, \bibinfo {author} {\bibfnamefont {F.}~\bibnamefont {Di~Mare}},
  \bibinfo {author} {\bibfnamefont {G.}~\bibnamefont {Nigro}}, \ and\ \bibinfo
  {author} {\bibfnamefont {L.}~\bibnamefont {Sorriso-Valvo}},\ }\href@noop {}
  {\bibfield  {journal} {\bibinfo  {journal} {Phys. Rev. E}\ }\textbf {\bibinfo
  {volume} {94}},\ \bibinfo {pages} {053109} (\bibinfo {year}
  {2016})}\BibitemShut {NoStop}%
\bibitem [{\citenamefont {Machicoane}\ \emph {et~al.}(2019)\citenamefont
  {Machicoane}, \citenamefont {Huck}, \citenamefont {Clark}, \citenamefont
  {Aliseda}, \citenamefont {Volk},\ and\ \citenamefont
  {Bourgoin}}]{Machicoane2019}%
  \BibitemOpen
  \bibfield  {author} {\bibinfo {author} {\bibfnamefont {N.}~\bibnamefont
  {Machicoane}}, \bibinfo {author} {\bibfnamefont {P.~D.}\ \bibnamefont
  {Huck}}, \bibinfo {author} {\bibfnamefont {A.}~\bibnamefont {Clark}},
  \bibinfo {author} {\bibfnamefont {A.}~\bibnamefont {Aliseda}}, \bibinfo
  {author} {\bibfnamefont {R.}~\bibnamefont {Volk}}, \ and\ \bibinfo {author}
  {\bibfnamefont {M.}~\bibnamefont {Bourgoin}},\ }in\ \href@noop {} {\emph
  {\bibinfo {booktitle} {Flowing Matter}}}\ (\bibinfo  {publisher} {Springer},\
  \bibinfo {year} {2019})\ pp.\ \bibinfo {pages} {177--209}\BibitemShut
  {NoStop}%
\bibitem [{\citenamefont {Boldyrev}(2005)}]{Boldyrev2005}%
  \BibitemOpen
  \bibfield  {author} {\bibinfo {author} {\bibfnamefont {S.}~\bibnamefont
  {Boldyrev}},\ }\href@noop {} {\bibfield  {journal} {\bibinfo  {journal}
  {Astrophys. J. Lett.}\ }\textbf {\bibinfo {volume} {626}},\ \bibinfo {pages}
  {L37} (\bibinfo {year} {2005})}\BibitemShut {NoStop}%
\bibitem [{\citenamefont {Brenner}\ \emph {et~al.}(2019)\citenamefont
  {Brenner}, \citenamefont {Eldredge},\ and\ \citenamefont
  {Freund}}]{brenner2019perspective}%
  \BibitemOpen
  \bibfield  {author} {\bibinfo {author} {\bibfnamefont {M.}~\bibnamefont
  {Brenner}}, \bibinfo {author} {\bibfnamefont {J.}~\bibnamefont {Eldredge}}, \
  and\ \bibinfo {author} {\bibfnamefont {J.}~\bibnamefont {Freund}},\
  }\href@noop {} {\bibfield  {journal} {\bibinfo  {journal} {Phys. Rev.
  Fluids}\ }\textbf {\bibinfo {volume} {4}},\ \bibinfo {pages} {100501}
  (\bibinfo {year} {2019})}\BibitemShut {NoStop}%
\bibitem [{\citenamefont {Jia}\ and\ \citenamefont
  {Ma}(2017)}]{doi:10.1190/geo2016-0300.1}%
  \BibitemOpen
  \bibfield  {author} {\bibinfo {author} {\bibfnamefont {Y.}~\bibnamefont
  {Jia}}\ and\ \bibinfo {author} {\bibfnamefont {J.}~\bibnamefont {Ma}},\
  }\href@noop {} {\bibfield  {journal} {\bibinfo  {journal} {Geophysics}\
  }\textbf {\bibinfo {volume} {82}},\ \bibinfo {pages} {V163} (\bibinfo {year}
  {2017})}\BibitemShut {NoStop}%
\bibitem [{\citenamefont {Appelhans}\ \emph {et~al.}(2015)\citenamefont
  {Appelhans}, \citenamefont {Mwangomo}, \citenamefont {Hardy}, \citenamefont
  {Hemp},\ and\ \citenamefont {Nauss}}]{APPELHANS201591}%
  \BibitemOpen
  \bibfield  {author} {\bibinfo {author} {\bibfnamefont {T.}~\bibnamefont
  {Appelhans}}, \bibinfo {author} {\bibfnamefont {E.}~\bibnamefont {Mwangomo}},
  \bibinfo {author} {\bibfnamefont {D.~R.}\ \bibnamefont {Hardy}}, \bibinfo
  {author} {\bibfnamefont {A.}~\bibnamefont {Hemp}}, \ and\ \bibinfo {author}
  {\bibfnamefont {T.}~\bibnamefont {Nauss}},\ }\href@noop {} {\bibfield
  {journal} {\bibinfo  {journal} {Spatial Statistics}\ }\textbf {\bibinfo
  {volume} {14}},\ \bibinfo {pages} {91 } (\bibinfo {year} {2015})}\BibitemShut
  {NoStop}%
\bibitem [{\citenamefont {Rasmussen}(2003)}]{rasmussen2003gaussian}%
  \BibitemOpen
  \bibfield  {author} {\bibinfo {author} {\bibfnamefont {C.~E.}\ \bibnamefont
  {Rasmussen}},\ }in\ \href@noop {} {\emph {\bibinfo {booktitle} {Summer School
  on Machine Learning}}}\ (\bibinfo {organization} {Springer},\ \bibinfo {year}
  {2003})\ pp.\ \bibinfo {pages} {63--71}\BibitemShut {NoStop}%
\bibitem [{\citenamefont {Lengyel}\ and\ \citenamefont
  {Friedrich}(2020)}]{Lengyel2020}%
  \BibitemOpen
  \bibfield  {author} {\bibinfo {author} {\bibfnamefont {J.}~\bibnamefont
  {Lengyel}}\ and\ \bibinfo {author} {\bibfnamefont {J.}~\bibnamefont
  {Friedrich}},\ }\enquote {\bibinfo {title} {Multiscale urban modeling},}\ in\
  \href {\doibase 10.1007/978-3-658-29746-6_32} {\emph {\bibinfo {booktitle}
  {Neue Dimensionen der Mobilit{\"a}t: Technische und betriebswirtschaftliche
  Aspekte}}},\ \bibinfo {editor} {edited by\ \bibinfo {editor} {\bibfnamefont
  {H.}~\bibnamefont {Proff}}}\ (\bibinfo  {publisher} {Springer Fachmedien
  Wiesbaden},\ \bibinfo {address} {Wiesbaden},\ \bibinfo {year} {2020})\ pp.\
  \bibinfo {pages} {387--408}\BibitemShut {NoStop}%
\end{thebibliography}%


\begin{thebibliography}{16}%
\makeatletter
\providecommand \@ifxundefined [1]{%
 \@ifx{#1\undefined}
}%
\providecommand \@ifnum [1]{%
 \ifnum #1\expandafter \@firstoftwo
 \else \expandafter \@secondoftwo
 \fi
}%
\providecommand \@ifx [1]{%
 \ifx #1\expandafter \@firstoftwo
 \else \expandafter \@secondoftwo
 \fi
}%
\providecommand \natexlab [1]{#1}%
\providecommand \enquote  [1]{``#1''}%
\providecommand \bibnamefont  [1]{#1}%
\providecommand \bibfnamefont [1]{#1}%
\providecommand \citenamefont [1]{#1}%
\providecommand \href@noop [0]{\@secondoftwo}%
\providecommand \href [0]{\begingroup \@sanitize@url \@href}%
\providecommand \@href[1]{\@@startlink{#1}\@@href}%
\providecommand \@@href[1]{\endgroup#1\@@endlink}%
\providecommand \@sanitize@url [0]{\catcode `\\12\catcode `\$12\catcode
  `\&12\catcode `\#12\catcode `\^12\catcode `\_12\catcode `\%12\relax}%
\providecommand \@@startlink[1]{}%
\providecommand \@@endlink[0]{}%
\providecommand \url  [0]{\begingroup\@sanitize@url \@url }%
\providecommand \@url [1]{\endgroup\@href {#1}{\urlprefix }}%
\providecommand \urlprefix  [0]{URL }%
\providecommand \Eprint [0]{\href }%
\providecommand \doibase [0]{http://dx.doi.org/}%
\providecommand \selectlanguage [0]{\@gobble}%
\providecommand \bibinfo  [0]{\@secondoftwo}%
\providecommand \bibfield  [0]{\@secondoftwo}%
\providecommand \translation [1]{[#1]}%
\providecommand \BibitemOpen [0]{}%
\providecommand \bibitemStop [0]{}%
\providecommand \bibitemNoStop [0]{.\EOS\space}%
\providecommand \EOS [0]{\spacefactor3000\relax}%
\providecommand \BibitemShut  [1]{\csname bibitem#1\endcsname}%
\let\auto@bib@innerbib\@empty
\bibitem [{\citenamefont {L{\'e}vy}\ and\ \citenamefont
  {Loeve}(1965)}]{levy1965processus}%
  \BibitemOpen
  \bibfield  {author} {\bibinfo {author} {\bibfnamefont {P.}~\bibnamefont
  {L{\'e}vy}}\ and\ \bibinfo {author} {\bibfnamefont {M.}~\bibnamefont
  {Loeve}},\ }\href@noop {} {\emph {\bibinfo {title} {Processus stochastiques
  et mouvement brownien}}}\ (\bibinfo  {publisher} {Gauthier-Villars Paris},\
  \bibinfo {year} {1965})\BibitemShut {NoStop}%
\bibitem [{\citenamefont {Lim}\ and\ \citenamefont {Muniandy}(2002)}]{lim2002}%
  \BibitemOpen
  \bibfield  {author} {\bibinfo {author} {\bibfnamefont {S.~C.}\ \bibnamefont
  {Lim}}\ and\ \bibinfo {author} {\bibfnamefont {S.~V.}\ \bibnamefont
  {Muniandy}},\ }\href {\doibase 10.1103/PhysRevE.66.021114} {\bibfield
  {journal} {\bibinfo  {journal} {Phys. Rev. E}\ }\textbf {\bibinfo {volume}
  {66}},\ \bibinfo {pages} {021114} (\bibinfo {year} {2002})}\BibitemShut
  {NoStop}%
\bibitem [{\citenamefont {Mandelbrot}\ and\ \citenamefont
  {Ness}(1968)}]{mandelbrot1968}%
  \BibitemOpen
  \bibfield  {author} {\bibinfo {author} {\bibfnamefont {B.~B.}\ \bibnamefont
  {Mandelbrot}}\ and\ \bibinfo {author} {\bibfnamefont {J.~W.~V.}\ \bibnamefont
  {Ness}},\ }\href {http://www.jstor.org/stable/2027184} {\bibfield  {journal}
  {\bibinfo  {journal} {SIAM Review}\ }\textbf {\bibinfo {volume} {10}},\
  \bibinfo {pages} {422} (\bibinfo {year} {1968})}\BibitemShut {NoStop}%
\bibitem [{\citenamefont {Grebenkov}\ \emph {et~al.}(2015)\citenamefont
  {Grebenkov}, \citenamefont {Belyaev},\ and\ \citenamefont
  {Jones}}]{grebenkov2015multiscale}%
  \BibitemOpen
  \bibfield  {author} {\bibinfo {author} {\bibfnamefont {D.~S.}\ \bibnamefont
  {Grebenkov}}, \bibinfo {author} {\bibfnamefont {D.}~\bibnamefont {Belyaev}},
  \ and\ \bibinfo {author} {\bibfnamefont {P.~W.}\ \bibnamefont {Jones}},\
  }\href@noop {} {\bibfield  {journal} {\bibinfo  {journal} {J. Phys. A}\
  }\textbf {\bibinfo {volume} {49}},\ \bibinfo {pages} {043001} (\bibinfo
  {year} {2015})}\BibitemShut {NoStop}%
\bibitem [{\citenamefont {Beran}(1994)}]{beran1994statistics}%
  \BibitemOpen
  \bibfield  {author} {\bibinfo {author} {\bibfnamefont {J.}~\bibnamefont
  {Beran}},\ }\href@noop {} {\emph {\bibinfo {title} {Statistics for
  long-memory processes}}},\ Vol.~\bibinfo {volume} {61}\ (\bibinfo
  {publisher} {CRC press},\ \bibinfo {year} {1994})\BibitemShut {NoStop}%
\bibitem [{\citenamefont {Flandrin}(1989)}]{flandrin1989spectrum}%
  \BibitemOpen
  \bibfield  {author} {\bibinfo {author} {\bibfnamefont {P.}~\bibnamefont
  {Flandrin}},\ }\href@noop {} {\bibfield  {journal} {\bibinfo  {journal} {IEEE
  Transactions on information theory}\ }\textbf {\bibinfo {volume} {35}},\
  \bibinfo {pages} {197} (\bibinfo {year} {1989})}\BibitemShut {NoStop}%
\bibitem [{\citenamefont {Davies}\ and\ \citenamefont
  {Harte}(1987)}]{Davies1987}%
  \BibitemOpen
  \bibfield  {author} {\bibinfo {author} {\bibfnamefont {R.~B.}\ \bibnamefont
  {Davies}}\ and\ \bibinfo {author} {\bibfnamefont {D.~S.}\ \bibnamefont
  {Harte}},\ }\href@noop {} {\bibfield  {journal} {\bibinfo  {journal}
  {Biometrika}\ }\textbf {\bibinfo {volume} {74}},\ \bibinfo {pages} {95}
  (\bibinfo {year} {1987})}\BibitemShut {NoStop}%
\bibitem [{fbm()}]{fbm}%
  \BibitemOpen
  \href@noop {} {}\bibinfo {howpublished}
  {https://pypi.org/project/fbm/}\BibitemShut {NoStop}%
\bibitem [{\citenamefont {Risken}(1996)}]{Risken}%
  \BibitemOpen
  \bibfield  {author} {\bibinfo {author} {\bibfnamefont {H.}~\bibnamefont
  {Risken}},\ }\href@noop {} {\emph {\bibinfo {title} {{The Fokker-Planck
  Equation}}}}\ (\bibinfo  {publisher} {Springer, Berlin},\ \bibinfo {year}
  {1996})\BibitemShut {NoStop}%
\bibitem [{\citenamefont {Mazzolo}(2017)}]{Mazzolo_2017}%
  \BibitemOpen
  \bibfield  {author} {\bibinfo {author} {\bibfnamefont {A.}~\bibnamefont
  {Mazzolo}},\ }\href {\doibase 10.1063/1.5000077} {\bibfield  {journal}
  {\bibinfo  {journal} {J. Math. Phys.}\ }\textbf {\bibinfo {volume} {58}},\
  \bibinfo {pages} {093302} (\bibinfo {year} {2017})}\BibitemShut {NoStop}%
\bibitem [{\citenamefont {Monin}\ and\ \citenamefont {Yaglom}(2007)}]{monin}%
  \BibitemOpen
  \bibfield  {author} {\bibinfo {author} {\bibfnamefont {A.~S.}\ \bibnamefont
  {Monin}}\ and\ \bibinfo {author} {\bibfnamefont {A.~M.}\ \bibnamefont
  {Yaglom}},\ }\href@noop {} {\emph {\bibinfo {title} {{Statistical Fluid
  Mechanics: Mechanics of Turbulence}}}}\ (\bibinfo  {publisher} {Courier Dover
  Publications},\ \bibinfo {year} {2007})\BibitemShut {NoStop}%
\bibitem [{\citenamefont {Stevens}(1995)}]{stevens1995six}%
  \BibitemOpen
  \bibfield  {author} {\bibinfo {author} {\bibfnamefont {C.~F.}\ \bibnamefont
  {Stevens}},\ }\href@noop {} {\emph {\bibinfo {title} {The six core theories
  of modern physics}}}\ (\bibinfo  {publisher} {MIT Press},\ \bibinfo {year}
  {1995})\BibitemShut {NoStop}%
\bibitem [{\citenamefont {Kolmogorov}(1962)}]{Kolmogorov1962}%
  \BibitemOpen
  \bibfield  {author} {\bibinfo {author} {\bibfnamefont {A.~N.}\ \bibnamefont
  {Kolmogorov}},\ }\href {\doibase 10.1017/S0022112062000518} {\bibfield
  {journal} {\bibinfo  {journal} {J. Fluid Mech.}\ }\textbf {\bibinfo {volume}
  {13}},\ \bibinfo {pages} {82} (\bibinfo {year} {1962})}\BibitemShut {NoStop}%
\bibitem [{\citenamefont {Oboukhov}(1962)}]{Oboukhov1962}%
  \BibitemOpen
  \bibfield  {author} {\bibinfo {author} {\bibfnamefont {A.~M.}\ \bibnamefont
  {Oboukhov}},\ }\href {\doibase 10.1017/S0022112062000506} {\bibfield
  {journal} {\bibinfo  {journal} {J. Fluid Mech.}\ }\textbf {\bibinfo {volume}
  {67}},\ \bibinfo {pages} {77} (\bibinfo {year} {1962})}\BibitemShut {NoStop}%
\bibitem [{\citenamefont {Friedrich}\ and\ \citenamefont
  {Peinke}(1997)}]{Friedrich:1997aa}%
  \BibitemOpen
  \bibfield  {author} {\bibinfo {author} {\bibfnamefont {R.}~\bibnamefont
  {Friedrich}}\ and\ \bibinfo {author} {\bibfnamefont {J.}~\bibnamefont
  {Peinke}},\ }\href@noop {} {\bibfield  {journal} {\bibinfo  {journal} {Phys.
  Rev. Lett.}\ }\textbf {\bibinfo {volume} {78}},\ \bibinfo {pages} {863}
  (\bibinfo {year} {1997})}\BibitemShut {NoStop}%
\bibitem [{\citenamefont {Yakhot}(2006)}]{yakhot:2006}%
  \BibitemOpen
  \bibfield  {author} {\bibinfo {author} {\bibfnamefont {V.}~\bibnamefont
  {Yakhot}},\ }\href@noop {} {\bibfield  {journal} {\bibinfo  {journal} {Phys.
  D}\ }\textbf {\bibinfo {volume} {215}},\ \bibinfo {pages} {166} (\bibinfo
  {year} {2006})}\BibitemShut {NoStop}%
\end{thebibliography}%
\end{document}


\title{Supplemental Material for ``Stochastic interpolation of sparsely sampled time series via multi-point fractional Brownian bridges''}
\author{J.~Friedrich}
\affiliation{Univ. Lyon, ENS de Lyon, Univ. Claude Bernard, CNRS, Laboratoire de Physique,
F-69342, Lyon, France}
\author{S.~Gallon}
\affiliation{Univ. Lyon, ENS de Lyon, Univ. Claude Bernard, CNRS, Laboratoire de Physique,
F-69342, Lyon, France}
\affiliation{Institute for Theoretical Physics I, Ruhr-University Bochum, Universit\"{a}tsstr. 150,
D-44801 Bochum, Germany}
\author{A.~Pumir}
\affiliation{Univ. Lyon, ENS de Lyon, Univ. Claude Bernard, CNRS, Laboratoire de Physique,
F-69342, Lyon, France}
\author{R.~Grauer}
\affiliation{Institute for Theoretical Physics I, Ruhr-University Bochum, Universit\"{a}tsstr. 150,
D-44801 Bochum, Germany}
\date{\today}
\maketitle

\section{Representations of fractional Brownian motion}
The literature on fractional Brownian motion typically distinguishes between two different representations of fBm. In the original work of L\'evy~\cite{levy1965processus}, fractional Brownian motion was introduced in form of a Riemann-Liouville integral with respect to a Gaussian white noise measure $\textrm{d}W(t)$
\begin{equation}
  X_{RL}(t)= \frac{1}{\Gamma\left(H+\frac{1}{2} \right)} \int_0^t
  \frac{1}{(t-t')^{\frac{1}{2}-H}}\textrm{d}W(t')\;,
  \label{eq:levy_fbm}
\end{equation}
where $\Gamma(z)$ denotes the Gamma function and $H<0<1$ the Hurst exponent.
This corresponds to a self-similar Gaussian stochastic process with zero mean whose increments, however, are nonstationary. Furthermore, its covariance
possesses a rather complicated functional form~\cite{lim2002}.
Therefore, Mandelbrot and van Ness~\cite{mandelbrot1968} (see also~\cite{grebenkov2015multiscale}) suggested a representation of fBm which is based on the Weyl integral
\begin{equation}
  X_{W}(t)= \frac{1}{\Gamma\left(H+\frac{1}{2} \right)} \left\{ \int_{-\infty}^0
  \left[\frac{1}{(t-t')^{\frac{1}{2}-H}} - \frac{1}{(-t')^{\frac{1}{2}-H}} \right]\textrm{d}W(t') + \int_0^t \frac{1}{(t-t')^{\frac{1}{2}-H}}\textrm{d}W(t')\right\}\;,
\end{equation}
for $t>0$. By contrast to the Riemann-Liouville representation (\ref{eq:levy_fbm}), this stochastic process possesses stationary increments \textrm{and} implies a simpler structure for the covariance
\begin{equation}
 \langle X_{W}(t)X_{W}(t') \rangle = \frac{\Gamma\left(1-2H \right) \cos(H\pi)}{2H\pi} \left(|t|^{2H}+|t'|^{2H}-|t-t'|^{2H}\right)\;,
\end{equation}
which was used throughout the constructions of fBm bridge processes in the main paper. Furthermore, as opposed to fractional Gaussian noise which exhibits a clear spectral density $\sim f^{1-2H}$~(we also refer the reader to the monograph~\cite{beran1994statistics}), here, the notion of spectral density has to be extended in order to capture the nonstationarity of fBm~\cite{flandrin1989spectrum}.

In addition, sample paths of fractional Brownian motion were constructed on the basis of the Davies-Harte algorithm~\cite{Davies1987}, which is similar to a Cholesky decomposition of the incremental covariance matrix.
The decomposition in the Davies-Harte algorithm~\cite{Davies1987} is achieved by embedding the original covariance matrix into a circulant matrix, which can readily be decomposed. Numerical implementations of this algorithm are available in form of Python packages~\cite{fbm}. The circulant embedding has the advantage that it is an exact numerical implementation of fBm, by contrast to other methods, e.g., Fourier representations of the integral (\ref{eq:levy_fbm}) which require a periodization of the kernels.

Due to the nonstationarity of the increments of fBm discussed above, some time series are also modeled in terms of Ornstein-Uhlenbeck processes due to its stationarity and mean-reverting properties~\cite{Risken}.
In the present study, however, we purely focused on the scaling range of the time series (i.e., the inertial range for the turbulence time signal). Therefore, no requirements for the saturation of the second order structure function at large scales (e.g., in Fig.1(b) in the main paper) had to be imposed. Consequently, we could rely on this rather simple method of the multi-point bridge construction on the basis of fBm. Nonetheless, a similar method that allows to constrain ordinary Ornstein-Uhlenbeck processes has been discussed recently~\cite{Mazzolo_2017}. It will be a task for the future to extend this treatment and introduce constrained fractional Ornstein-Uhlenbeck processes.

Whereas the use of Ornstein-Uhlenbeck processes appears as
a natural way to take into account the saturation of the structure function at
large scales, other methods need to be developed to account for the existence
of ranges of scales, where the process is known to develop other power laws,
as it occurs in the dissipative range in turbulent flows~\cite{monin}.

\section{Calculation of conditional moments for centered Gaussian process}
\label{app:gauss}
Although the conditional moments of a multivariate Gaussian distribution are
well known, in what follows, we want to give a brief derivation. To this end, we consider the characteristic functional
\begin{equation}
  \varphi[\alpha]= \left \langle e^{i \int \textrm{d}t \alpha(t)X(t)} \right \rangle\;.
\end{equation}
For a centered Gaussian process $X(t)$, the characteristic functional reduces to (see, for instance~\cite{monin,stevens1995six})
\begin{equation}
  \varphi[\alpha] = e^{-\frac{1}{2}\int \textrm{d}t \int  \textrm{d}t' \alpha(t) \left \langle X(t)X(t') \right \rangle \alpha(t')}\;.
\end{equation}
The $n$-point probability density function (PDF)
\begin{equation}
f_n(X_1, t_1; \ldots; X_n ,t_n)=
\prod_{i=1}^n \langle \delta(X(t_i)-X_i) \rangle\;,
\end{equation}
can thus be re-expressed according to
\begin{align}\nonumber
  \lefteqn{f_n(X_1, t_1; \ldots; X_n ,t_n)
  = \int \frac{\textrm{d} k_1}{2 \pi} \ldots \frac{\textrm{d} k_n}{2 \pi}
  e^{-i \sum_{i=1}^n k_i X_i} \left \langle e^{i \sum_{i=1}^n k_i X(t_i)} \right \rangle} \\ \nonumber
  &= \int \frac{\textrm{d} k_1}{2 \pi} \ldots \frac{\textrm{d} k_n}{2 \pi}
  e^{-i \sum_{i=1}^nk_i X_i} \varphi \left[\alpha(t)= \sum_{i=1}^n k_i \delta(t-t_i) \right]\\
  &= \int \frac{\textrm{d} k_1}{2 \pi} \ldots \frac{\textrm{d} k_n}{2 \pi}
  e^{-i \sum_{i=1}^n k_i X_i} \exp \left [-\frac{1}{2}\sum_{i=1}^n \sum_{i'=1}^n
  k_i k_{i'} \left \langle X(t_i)X(t_{i'}) \right \rangle \right ]\;.
\end{align}
Therefore, we obtain a multivariate Gaussian distribution
\begin{equation}
f_n(X_1, t_1; \ldots; X_n ,t_n)
  =\frac{1}{\sqrt{(2 \pi)^n \det(\sigma)}} e^{-\frac{1}{2} \mathbf{ X}^T \sigma^{-1} \mathbf{ X}}\;,
\end{equation}
where
\begin{equation}
  \mathbf{X}=(X_1, \ldots, X_n)^T\;,
\end{equation}
and where
\begin{equation}
  \sigma_{ij} = \langle X(t_i)X(t_j)\rangle\;,
\end{equation}
denotes the covariance matrix. Next, we can calculate the $n+1$-point quantity,
\begin{align}\nonumber
  \lefteqn{\prod_{i=1}^n \langle X(t)\delta(X(t_i)-X_i) \rangle
  =\int \frac{\textrm{d} k_1}{2 \pi} \ldots \frac{\textrm{d} k_n}{2 \pi}
  e^{-i \sum_{i=1}^n k_i X_i} \left \langle X(t) e^{i \sum_{i=1}^n k_i X(t_i)} \right \rangle} \\ \nonumber
  &= \int \frac{\textrm{d} k_1}{2 \pi} \ldots \frac{\textrm{d} k_n}{2 \pi}
  e^{-i \sum_{i=1}^nk_i X_i} \left. \frac{ \delta \varphi[\alpha] }{\delta i \alpha(t)} \right |_{\alpha(t)=\sum_{i=1}^n k_i \delta(t-t_i)}\\ \nonumber
  &=  i \int \frac{\textrm{d} k_1}{2 \pi} \ldots \frac{\textrm{d} k_n}{2 \pi}
  e^{-i \sum_{i=1}^n k_i X_i}  \sum_{i=1}^n k_i \left \langle X(t)X(t_i) \right \rangle e^{-\frac{1}{2}\sum_{i=1}^n \sum_{i'=1}^n
  k_i k_{i'} \left \langle X(t_i)X(t_{i'}) \right \rangle}\\
  &= -\sum_{i=1}^n  \left \langle X(t)X(t_i) \right \rangle \frac{\textrm{d}}{\textrm{d}X_i} f_n(X_1, t_1; \ldots; X_n ,t_n)\;,
\end{align}
\newpage
as well as the $n+2$-point quantity
\begin{align}\nonumber
  \lefteqn{\prod_{i=1}^n \langle X(t)X(t')\delta(X(t_i)-X_i) \rangle=\int \frac{\textrm{d} k_1}{2 \pi} \ldots \frac{\textrm{d} k_n}{2 \pi}
  e^{-i \sum_{i=1}^n k_i X_i} \left \langle X(t)X(t') e^{i \sum_{i=1}^n k_i X(t_i)} \right \rangle} \\ \nonumber
  &= \int \frac{\textrm{d} k_1}{2 \pi} \ldots \frac{\textrm{d} k_n}{2 \pi}
  e^{-i \sum_{i=1}^nk_i X_i} \left. \frac{ \delta^2 \varphi[\alpha] }{\delta i \alpha(t)\delta i \alpha(t')} \right |_{\alpha(t)=\sum_{i=1}^n k_i \delta(t-t_i)}\\ \nonumber
  &=  \int \frac{\textrm{d} k_1}{2 \pi} \ldots \frac{\textrm{d} k_n}{2 \pi}
  e^{-i \sum_{i=1}^n k_i X_i} \left [\left \langle X(t)X(t') \right \rangle  -\sum_{i=1}^n \sum_{i'=1}^n k_i k_{i'}\left \langle X(t)X(t_i) \right \rangle
  \left \langle X(t')X(t_{i'}) \right \rangle \right]e^{-\frac{1}{2}\sum_{i=1}^n \sum_{i'=1}^n
  k_i k_{i'} \left \langle X(t_i)X(t_{i'}) \right \rangle }\\
  &= \left[ \left \langle X(t)X(t') \right \rangle +\sum_{i=1}^n \sum_{i'=1}^n   \left \langle X(t)X(t_i) \right \rangle
  \left \langle X(t')X(t_{i'}) \right \rangle \frac{\textrm{d}^2}{\textrm{d}X_i \textrm{d}X_{i'}} \right] f_n(X_1, t_1; \ldots; X_n ,t_n)\;.
  \;,
\end{align}
The $n$-times conditional moments
\begin{align}
  \label{eq:cond1}
  \langle X(t)|\{X_i, t_i\} \rangle =& \frac{\langle X(t) \prod_{i=1}^n \delta(X(t_i)-X_i) \rangle}{\prod_{i=1}^n \langle \delta(X(t_i)-X_i) \rangle}\;,\\
  \langle X(t)X(t')|\{X_i, t_i\}\rangle =& \frac{\langle X(t)X(t') \prod_{i=1}^n \delta(X(t_i)-X_i) \rangle}{\prod_{i=1}^n \langle \delta(X(t_i)-X_i) \rangle}\;,
  \label{eq:cond2}
\end{align}
thus read
\begin{equation}
  \langle X(t)|\{X_i, t_i\} \rangle =  \sigma_{ij}^{-1} \left \langle X(t)X(t_i) \right \rangle X_j \;,
  \label{eq:cond1_gauss}
\end{equation}
and
\begin{align}
\langle X(t)X(t')|\{X_i, t_i\}\rangle= \left \langle X(t)X(t') \right \rangle-\left[\sigma_{ij}^{-1} - \sigma_{ik}^{-1}X_k X_l\sigma_{jl}^{-1} \right] \left \langle X(t)X(t_i) \right \rangle
  \left \langle X(t')X(t_j) \right \rangle \;.
  \label{eq:cond2_gauss}
\end{align}
where we made use of the symmetry of the covariance matrix and where we imply summation over equal indices.
\section{Proof for moments of multi-point fractional Brownian bridge}
\label{app:fBb}
In this section, we want to proof that the multi-point fBb
\begin{equation}
  X^B(t) =X(t) -  (X(t_i)-X_i) \sigma_{ij}^{-1}
  \left \langle X(t)X(t_j) \right \rangle\;,
  \label{eq:gen_bridge}
\end{equation}
possesses identical one and two-points moments to the moments fBm process $X(t)$ \emph{conditioned on} $\{X_i, t_i\}$, i.e., eqs. (\ref{eq:cond1_gauss}-\ref{eq:cond2_gauss}).
Therefore, we first calculate the mean of the fBb in eq. (\ref{eq:gen_bridge})
\begin{align}\nonumber
  \left \langle X^B(t) \right \rangle&=
  \underbrace{\left \langle X(t) \right \rangle}_{=0}-
  ( \underbrace{\left \langle X(t) \right \rangle}_{=0} -X_i) \sigma_{jk}^{-1}
  \langle X(t)X(t_j)\rangle \\
  &= X_i \sigma_{jk}^{-1}
  \langle X(t)X(t_j)\rangle\;,
\end{align}
where we assumed that the fBm possesses zero mean. Next, the correlation function of the generalized fBb reads
\begin{align}\nonumber
    \lefteqn{\left \langle X^B(t) X^B(t') \right \rangle}\\ \nonumber
    &= \langle X(t)X(t') \rangle-
    2 \langle X(t)X(t_i) \rangle \sigma_{ij}^{-1} \langle X(t)X(t_j) \rangle +\underbrace{\langle X(t_k)X(t_l) \rangle}_{\sigma_{kl}} \sigma_{ik}^{-1} \langle X(t)X(t_i) \rangle \langle X(t')X(t_j) \rangle \sigma_{jl}^{-1}\\ \nonumber
    &~~+ \sigma_{ik}^{-1}X_k \sigma_{jl}^{-1} X_l \left \langle X(t)X(t_i) \right \rangle
    \left \langle X(t')X(t_j) \right \rangle\\
    &= \left \langle X(t)X(t') \right \rangle-\sigma_{ij}^{-1} \left \langle X(t)X(t_i) \right \rangle  \left \langle X(t')X(t_j) \right \rangle+ \sigma_{ik}^{-1}X_k \sigma_{jl}^{-1} X_l \left \langle X(t)X(t_i) \right \rangle
      \left \langle X(t')X(t_j) \right \rangle \;,
\end{align}
where we made use of the identity $\sigma_{kl}\sigma_{ki}^{-1}=\delta_{il}$.
\section{Error analysis of the optimal Hurst exponent determination}
\begin{figure}[h]
\centering
\includegraphics[width=0.32 \textwidth]{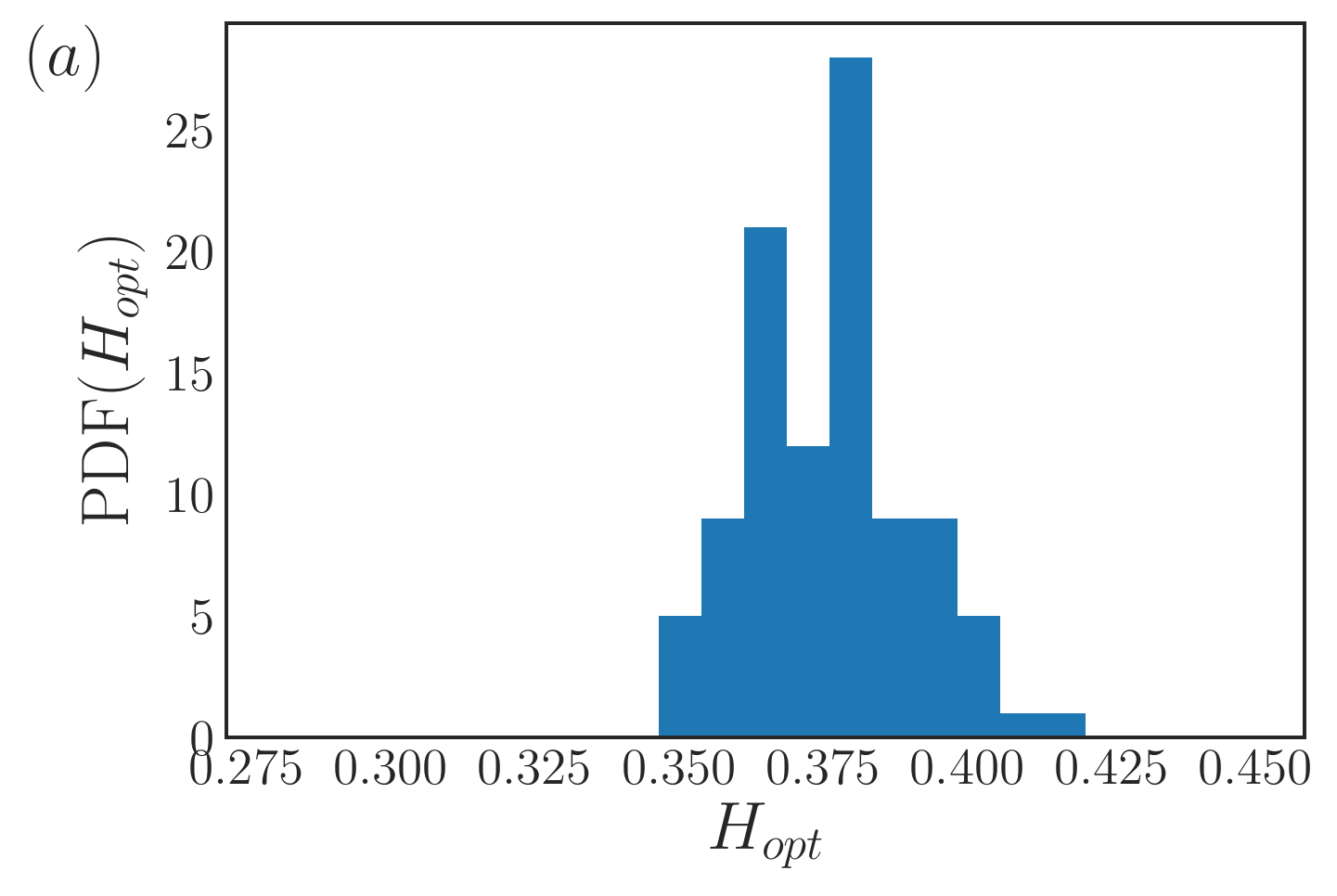}
\includegraphics[width=0.32 \textwidth]{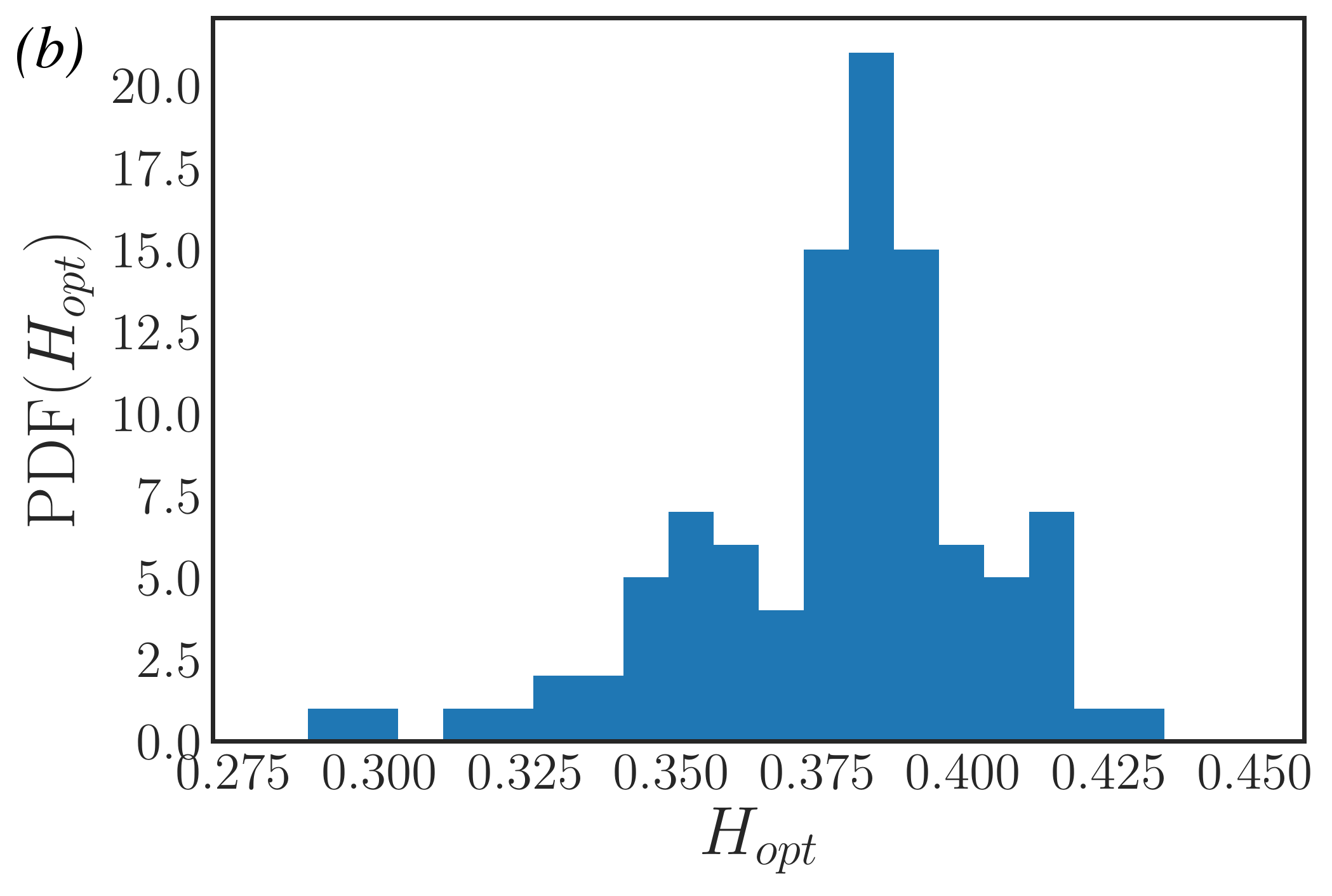}
\includegraphics[width=0.32 \textwidth]{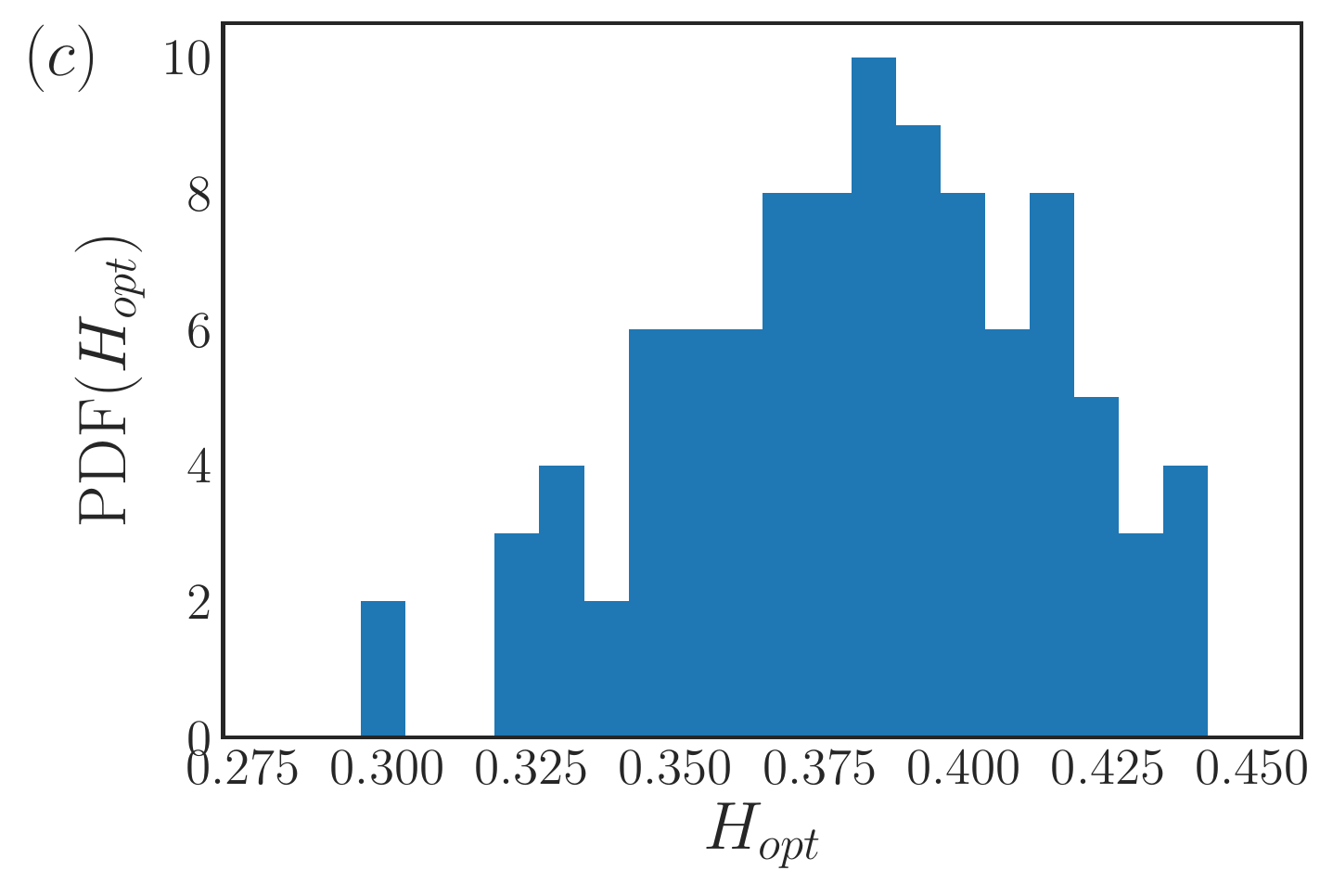}
\caption{(a) Histogram of the optimal Hurst exponent evaluated from 100 synthetic samples of fBm with $\tilde H=0.3786$. (b) Same plot as in (a), but now for 100 sub-samples of the randomized turbulent signal (Fig. 2(b) in main paper). (c) Same plot, but now for 100 sub-samples of the turbulent signal (Fig. 2(a) in main paper).}
\label{fig:hist}
\end{figure}

An optimization procedure similar to the one used to generate Fig. 1(c) of the main paper for different synthetic samples drawn as fBm with Hurst index $\tilde H$ yields slightly different results for the optimal Hurst exponent $H_{opt}$. In order to compare to the optimization procedure for the turbulent signal later, we choose 100 synthetic samples of fBm with Hurst exponent $\tilde H=0.3786$ with a spatial resolution of $\tilde N =128$ points. Fig.~\ref{fig:hist}(a) depicts the histograms for the optimal Hurst exponent obtained via optimization procedured as discussed in the main paper. The histogram is clearly peaked at the prescribed values $\tilde H$ and we obtain $\langle H_{opt} \rangle =0.3783 \pm 0.013$.
\begin{figure}[h]
\centering
\includegraphics[width=0.32 \textwidth]{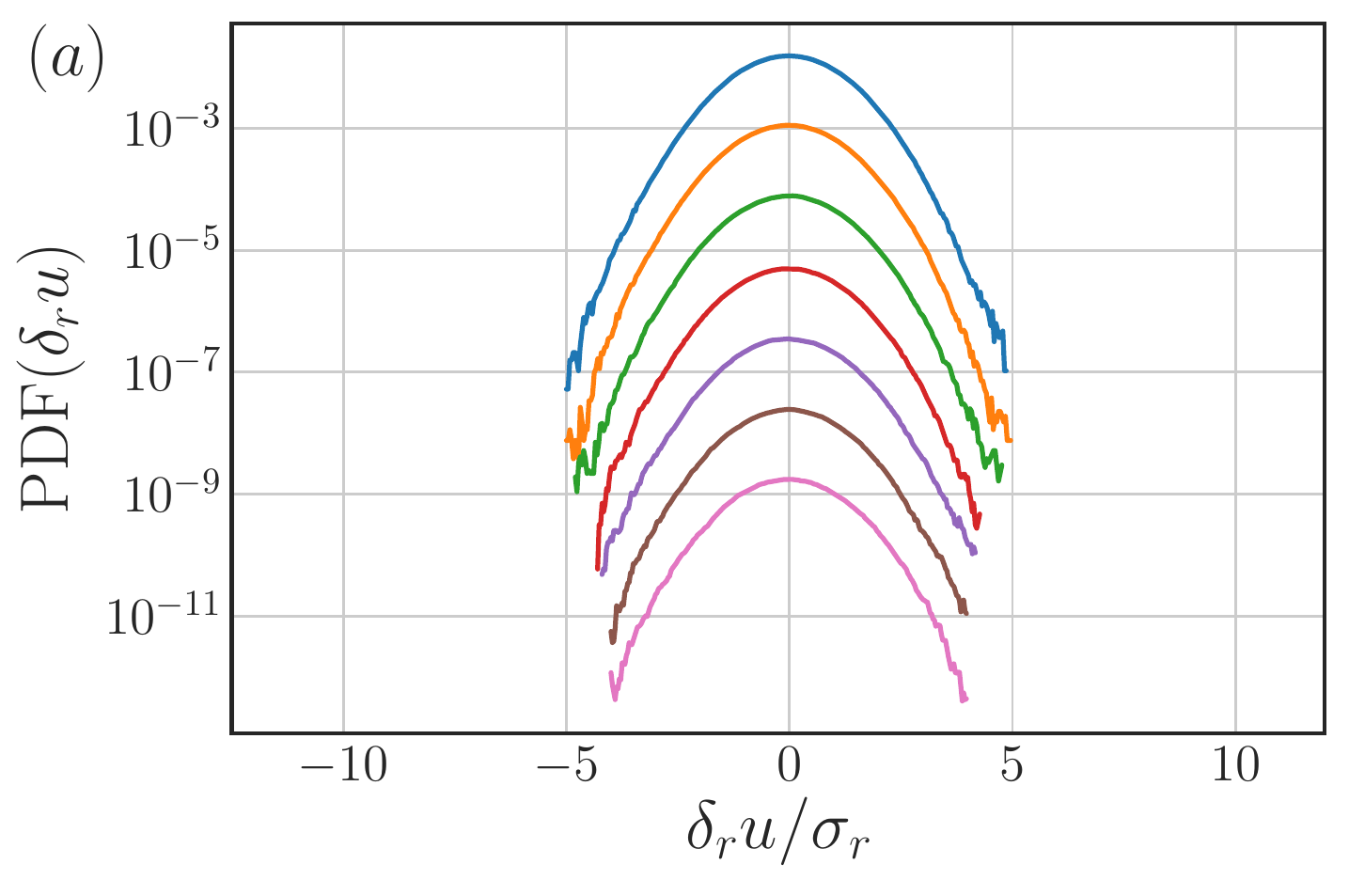}
\includegraphics[width=0.32 \textwidth]{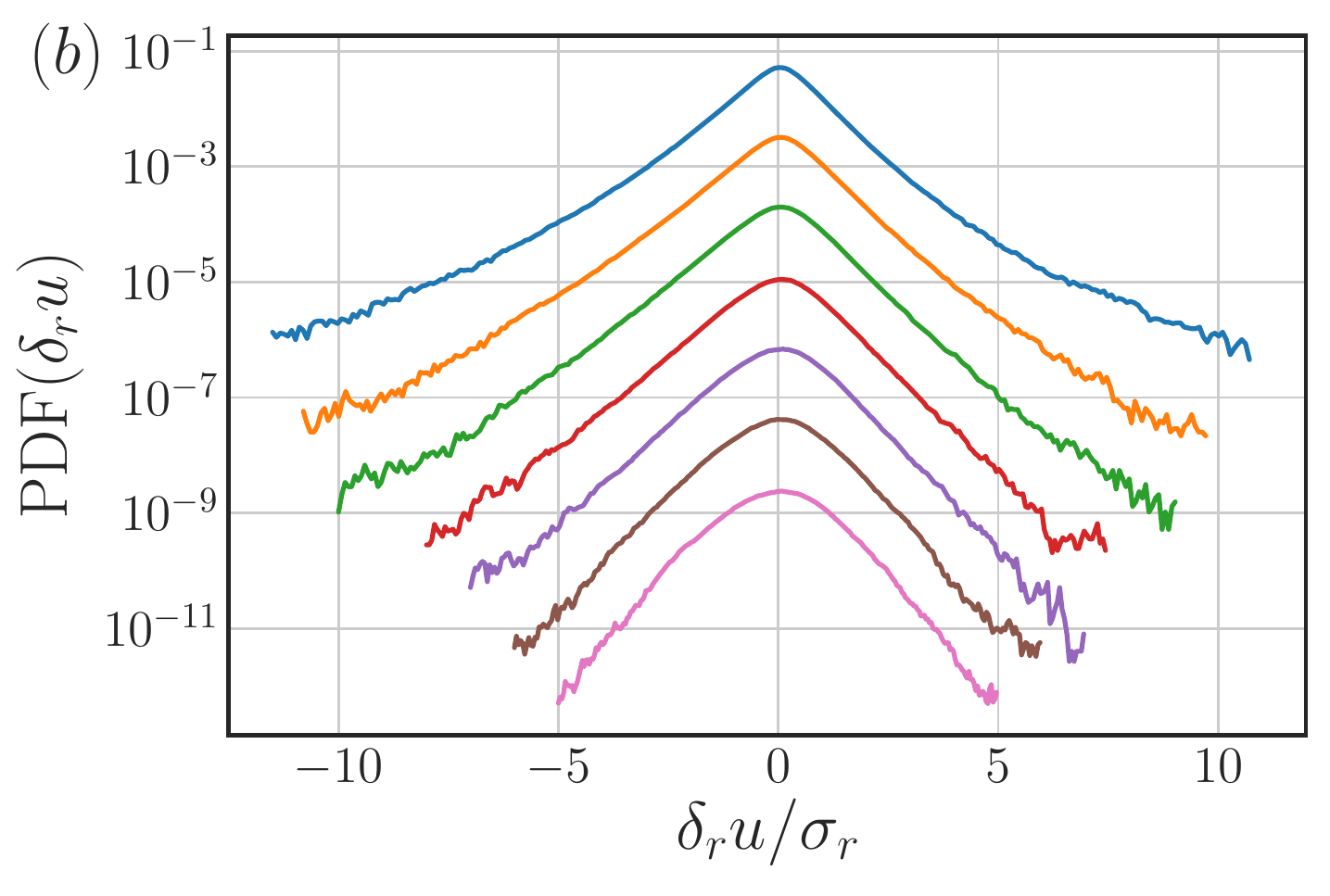}
\includegraphics[width=0.32 \textwidth]{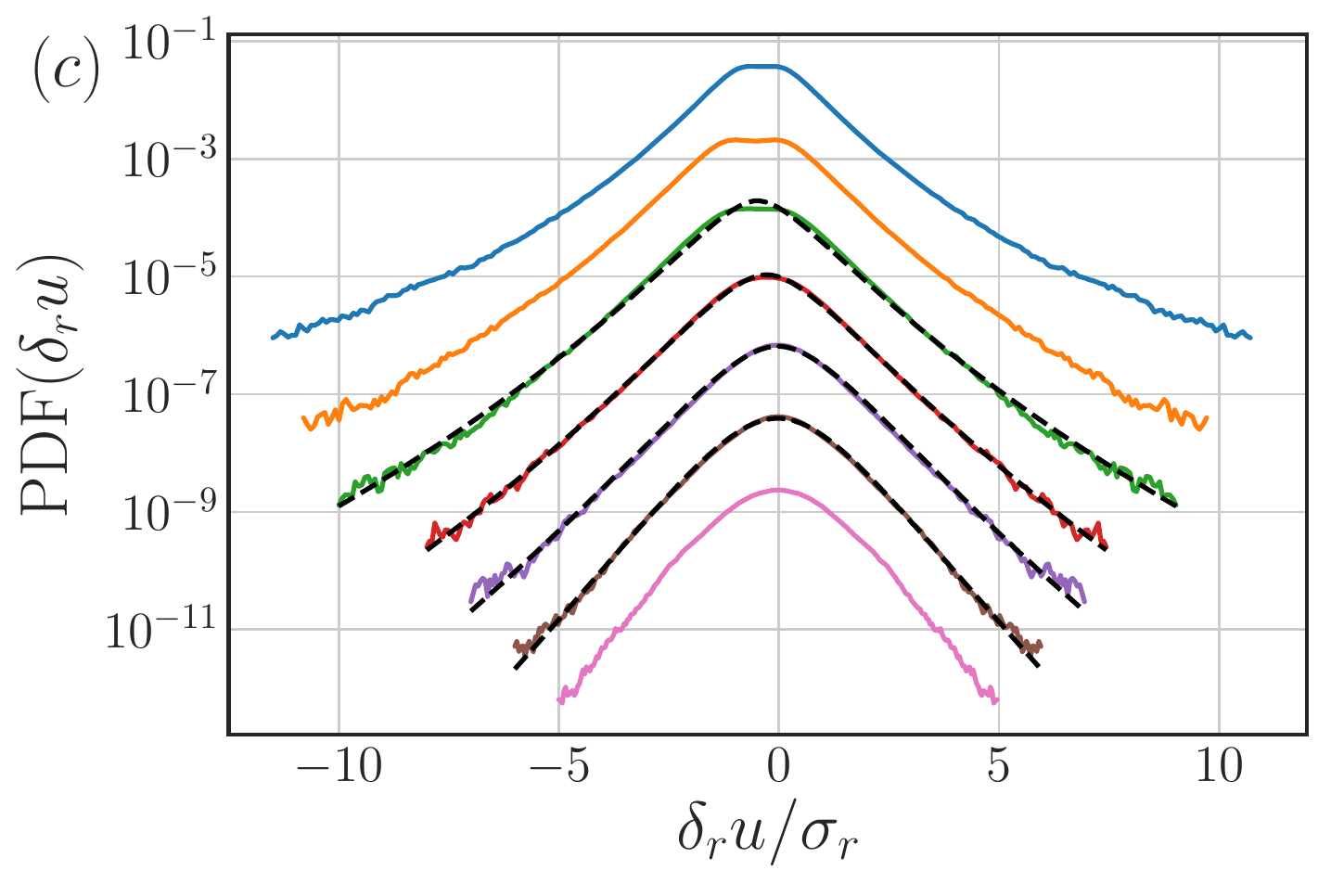}
\caption{(a) PDFs of velocity increments $\delta_r u$ ($\log(r/L)=   -6.41, -5.60, -4.66, -3.64, -2.75, -1.86, -0.87, 0.12$ from top to bottom) of the randomized turbulent signal which has been obtained by random rotation of Fourier phases from the original turbulent signal in (b). The PDFs in (a) exhibit self-similarity across different scales and the randomized signal can thus be used in the described optimization procedure via fBbs. (c) symmetrization of the PDFs in (b). The dashed curves correspond to the prediction of the K62 model of turbulence with $\mu= $ and are only shown for PDFs with scale separation $r$ in the inertial range.}
\label{fig:pdfs}
\end{figure}
These variations are due to finite-size effects as well as noise due to the numerical implementation of the Davies-Harte algorithm. Turning next to the turbulent signal in Fig. 2(a) of the main paper, we first perform a Fourier transform of the entire signal and then randomize Fourier phases in order to get rid of strong (intermittent) correlations. The resulting signal possesses Gaussian statistics which can be verified by computing the PDF of velocity increments $\delta_r u = u(x+r)-u(x)$ at different scales $r$, as it has been done in  Fig.~\ref{fig:pdfs}(a). For comparison, Fig.~\ref{fig:pdfs}(b) depicts the velocity increment PDF of the original turbulent signal for the same scale separations $r$ as in (a). As to be expected the PDFs develop pronounced tails at small scales, a key signature of small-scale intermittency. Due to the self-similar property of the randomized signal in Fig.~\ref{fig:pdfs}(a), the corresponding fluctuations can be treated by the same optimization routine as the synthetic fBm. However,
as explained in the main text, one has to take care of the existence of a dissipation scale in both the turbulent and randomized signal. To this end, we consider plateaus in log-derivatives of structure functions and determine the small-scale cut-off to be approximately $96 \eta$, where $\eta$ denotes the Kolmogorov microscale.
We thus perform the optimization procedure for 100 different sub-samples of the randomized signal.
The result is shown in Fig.~\ref{fig:hist}(b) and shows many similarities to the one obtained from synthetic samples in Fig.~\ref{fig:hist}(a), although the standard deviation is slightly higher. In fact, we obtain $\langle H_{opt} \rangle=0.3786 \pm 0.0251$.
\begin{figure}[h]
\centering
\includegraphics[width=0.4 \textwidth]{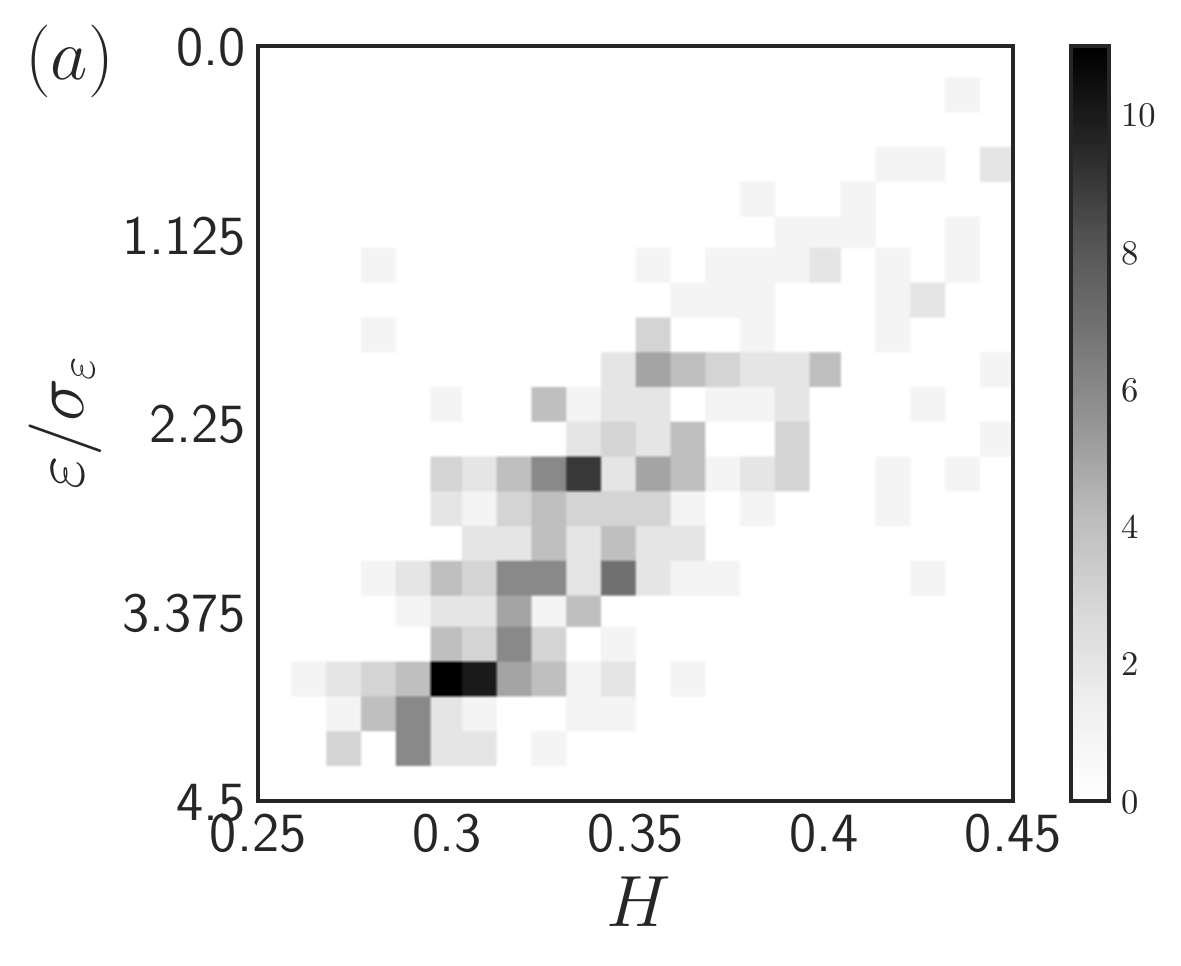}
\includegraphics[width=0.5 \textwidth]{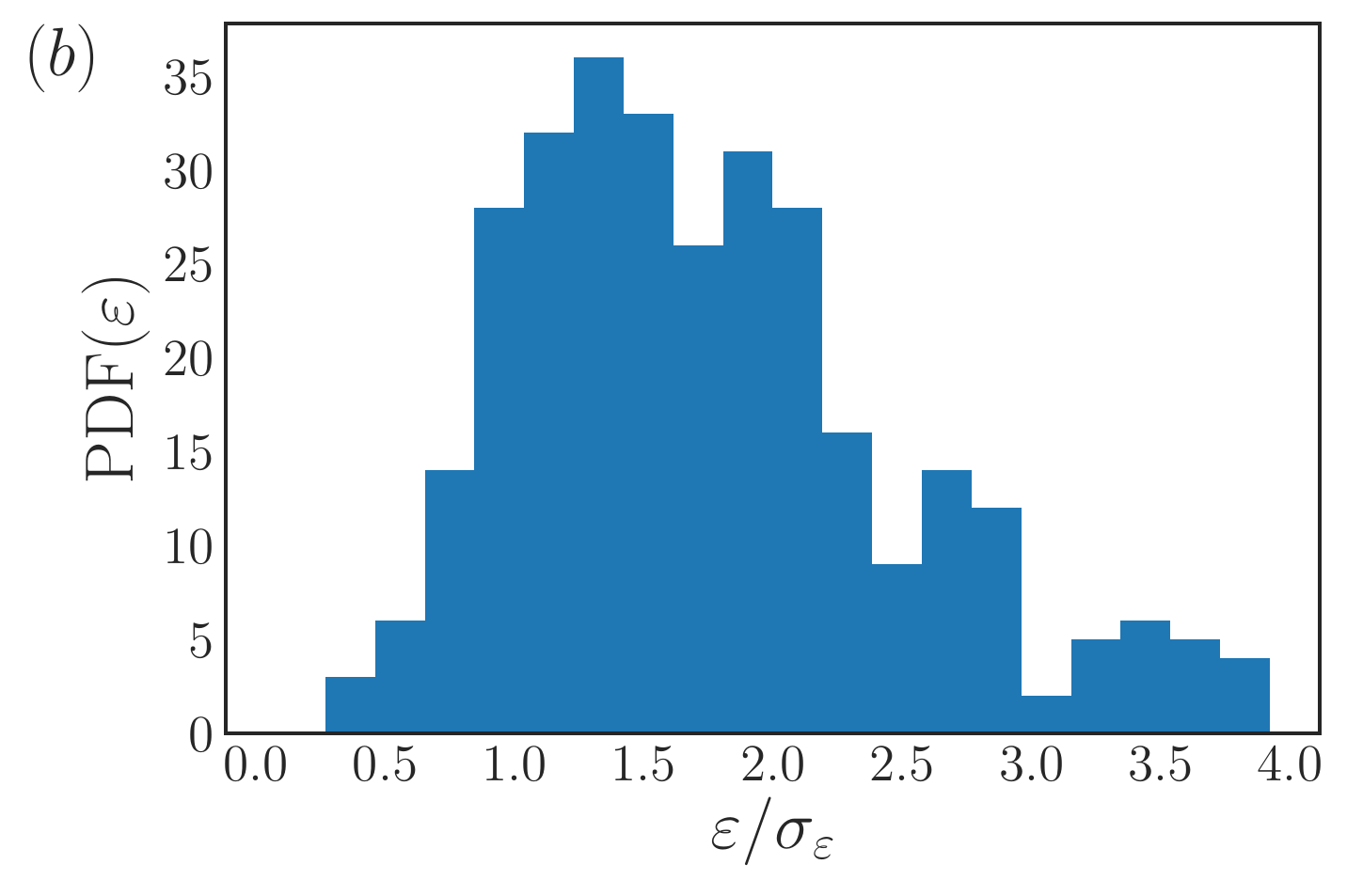}
\caption{(a) Joint histogram of energy dissipation rate and Hurst exponent evaluated from 320 different sub-samples of the turbulent signal. The axis of the energy dissipation has been divided by the standard deviation determined from all sub-samples. (b) Histogram of the energy dissipation rate.}
\label{fig:joint}
\end{figure}

In order to assess this result, we consider the K62 model of turbulence~\cite{Kolmogorov1962,Oboukhov1962}, which predicts structure functions
\begin{equation}
  \langle (\delta_r u)^n \rangle \sim |r|^{\zeta_n} \quad \textrm{with scaling exponents} \quad \zeta_n = \frac{n}{3}-\frac{\mu}{18} n(n-3)=an-bn^2\;.
\label{eq:k62}
\end{equation}
Latter model can also be cast as a stochastic process for the velocity increment $\delta_r u$ which evolves from large scales $L$ to smaller scales~\cite{Friedrich:1997aa}. More precisely, the Langevin equation with
\begin{equation}
  -\textrm{d}(\delta_r u) = -\frac{2+\mu}{6r} \delta_r u \textrm{d}r + \sqrt{\frac{\mu}{18r}}\textrm{d}W(r)\;,
\end{equation}
where $W(r)$ is white noise. Hence, intermittency in the lognormal model of turbulence is described in terms of a multiplicative noise term in the corresponding Langevin equation for the velocity increments. As, by virtue of the randomization of Fourier phases described above, we actually suppress this part of the signal, we are able to compare the remaining self-similar part in the K62 model, which
is given by $H_{K62}= (2+\mu)/6$ to our analysis.
The estimated value $\langle H_{opt} \rangle $ thus implies that $\mu = 0.2716 \pm 0.1506$. This result, which has been obtained from the optimization procedure of the randomized signal can now be compared to the original turbulent case as follows: First, the velocity increment PDF of the K62 model is accessible via a Mellin transform of the scaling exponents (\ref{eq:k62}), which yields~\cite{yakhot:2006}
\begin{equation}
    f(\delta_r u) = \frac{1}{2\pi \delta_r u\sqrt{\ln r^b}}
    \int_{-\infty}^{\infty} dx \; e^{-x^2} \exp\left[- \frac{\left(\ln \frac{\delta_r u}{r^a\sqrt{2}x}\right)^2}{4 b \ln r}\right]\;,
    \label{eq:pdfK62}
\end{equation}
This PDF is now used as a fit for the velocity increment PDFs in Fig.~\ref{fig:pdfs}(b) after a symmetrization (the PDFs in Fig.~\ref{fig:pdfs}(b) are strongly skewed, a direct consequence of the turbulent energy transfer). The symmetrization is depicted in Fig.~\ref{fig:pdfs}(c) and fitting the whole range of scales that were used for the optimization procedure from above, we obtain $\mu = 0.2913 \pm 0.0833$. Examples of the corresponding PDF (\ref{eq:pdfK62}) for different scales $r$ are indicated as the dashed curves in Fig.~\ref{fig:pdfs}(c). Hence, the intermittency coefficient determined from the optimization procedure of the randomized signal and the intermittency coefficient from the turbulent signal agree fairly well.

Finally, we applied the optimization procedure to the original turbulent signal in keeping the same range of scales and number of sub-samples as for the randomized signal. The result is shown in Fig.~\ref{fig:hist}(c). Compared to (a) and (b) the optimal Hurst exponent exhibits stronger sample-to-sample fluctuations. The mean value, however, agrees very well with the Hurst exponent of the randomized signal and we obtain $\langle H_{opt} \rangle= 0.3797 \pm 0.0368$. In order to further categorize these fluctuations, which can be considered as a direct signature of intermittency (i.e., non-self-similarity), we determined the joint PDF of the optimal Hurst parameter and the energy dissipation rate $\varepsilon = 2\nu  \left(\frac{\partial u(x)}{\partial x} \right)^2 $ averaged over each sample. The result is depicted in Fig.~\ref{fig:joint} (a). Even with limited statistics, we observe a clear negative correlation between energy dissipation and the determined value of the optimal Hurst exponent.
\bibliographystyle{apsrev4-1}
\bibliography{sm.bib}